\documentclass[showpacs, amsmath, amssymb, amsthm, aps, jmp, thmsb, twocolumn]{revtex4-1}

\usepackage{color,amsbsy,amssymb,latexsym,amsfonts, amsmath,cancel}
\usepackage{mathrsfs}
\usepackage{graphicx}
\usepackage{bbm}

\usepackage{comment}

\newcommand{\bel}[1]{\begin{equation}\label{#1}}                     
\newcommand{\bal}[1]{\begin{eqnarray}\label{#1}}   
\newcommand{\be}{\begin{equation}}               
\newcommand{\ba}{\begin{eqnarray}}           
\newcommand{\ee}{\end{equation}}
\newcommand{\ea}{\end{eqnarray}}

\newcommand{\bra}[1]{\left\langle #1 \right|}
\newcommand{\ket}[1]{\left| #1 \right\rangle}

\newcommand{\bea}{\begin{equation}}
\newcommand{\eea}{\end{equation}}

\begin{document}
%%%%%%%%    file name   %%%%%%%%%%%%%%%%%%%%
\title{Uncertainty of Weak Measurement and Merit of Amplification}

\author{Jaeha Lee}
\email[Email:]{jlee@post.kek.jp}
\affiliation{Department of Physics, University of Tokyo, 7-3-1 Hongo, Bunkyo-ku, Tokyo 113-0033, Japan}

\author{Izumi Tsutsui}
\email[Email:]{izumi.tsutsui@kek.jp}
\affiliation{Theory Center, Institute of Particle and Nuclear Studies,
High Energy Accelerator Research Organization (KEK), 1-1 Oho, Tsukuba, Ibaraki 305-0801, Japan}

\date{\today}

\begin{abstract}
Aharonov's weak value, which is a physical quantity obtainable by weak measurement, admits amplification and hence is deemed to be useful for precision measurement.
We examine the significance of  the amplification based on the uncertainty of measurement, and show that the trade-offs among the three  (systematic, statistical and nonlinear) components
of the uncertainty inherent in the weak measurement will set an upper limit on the usable amplification.  Apart from the Gaussian state models employed for demonstration, 
our argument is completely general; it is free from approximation and valid for arbitrary observables $A$ and couplings $g$.
\end{abstract}


\maketitle

%%%%%%%%%%%%%%%%%%%%%%%%%%%%%%
\phantom{}
\paragraph*{Introduction.}
%%%%%%%%%%%%%%%%%%%%%%%%%%%%%%

The novel physical quantity in quantum mechanics called {\it weak value}, proposed by Aharonov and co-workers long ago \cite{Aharonov_1964, Aharonov_1988}, has gained a renewed interest in recent years.   One of the reasons for this is  
that, unlike the standard physical value given by an eigenvalue of an observable $A$, the weak value may be considered meaningful even for a set of non-commutable observables, simultaneously.  This inspired a new insight for understanding counter-intuitive phenomena, such as the three-box paradox \cite{Aharonov_1991} and Hardy's paradox  \cite{Yokota_2009}.  The other, perhaps stronger, motive comes from the realization that the weak value can be amplified by adjusting the process of measurement, 
{\it weak measurement}.  Specifically, by choosing properly the initial and the final state (pre and postselection) of the process,  the weak value can be made arbitrarily large, and this may be utilized for precision measurement.  In fact, it has been reported that a significant amplification is achieved to observe successfully the spin Hall effect of light \cite{Hosten_2008}.  A similar amplification has also been shown to be available for detecting ultrasensitive beam deflection in a Sagnac interferometer \cite{Dixon_2009}.   

In view of this, it is quite natural to ask whether there exists a limit on amplification, and if so why.  This question was addressed recently in 
 \cite{Wu_2011}, which extended the treatment of \cite{Josza_2007} to the full order of the coupling $g$ between the system and the measurement device,  where the amplification is analyzed based on the average shift of the meter of the device.  For the particular case of the observable $A$ fulfilling $A^2 = \mathrm{Id}$ and the Gaussian device states, it has been shown that the amplification rate, as well as the signal-to-noise ratio, has an upper limit  \cite{Koike_2011, Nakamura_2012}.  No such limit arises if the device state can be tuned precisely according to the weak value and the coupling \cite{Susa_2012}.    
 
This Letter presents a completely new approach to the analysis of the weak value amplification.  Rather than focusing on the shift of the meter, we consider the uncertainty of the weak measurement and inquire when it is meaningful as measurement for given particular purposes.  This should be more important, because during the amplification the obtained amplified shift may well drift away from the intended weak value, rendering the whole measurement meaningless.   Our uncertainty is defined from a probabilistic estimation on the deviation of the measured value from the weak value itself, and is shown to be separable into systematic, statistical, and nonlinear components.   
For the purpose of probing the existence of a physical effect, as done in the experiments \cite{Hosten_2008, Dixon_2009},
the condition of
significant weak measurement is presented by demanding that the uncertainty (with the probability assigned beforehand) be smaller than the weak value to be measured.
In the case of the Gaussian state,  this explains the appearance of the range of amplification in which the existence of the effect is affirmed with assurance greater than the probability.   The case also confirms the anticipation \cite{Aharonov_2005} that 
the weak value may be observed with non-small $g$, {\it i.e.}, by \lq non-weak\rq\ measurement.
Apart from the Gaussian state analysis, our treatment is completely general; it is valid for an arbitrary dimensional system with arbitrary observables $A$ for all range of couplings $g$, and no approximation is used throughout (detailed discussions with mathematical proofs shall be given in Supplemental Material).

%%%%%%%%%%%%%%%%%%%%%%%%%%%%%%
\phantom{}
\paragraph*{Weak Value and Weak Measurement.}
%%%%%%%%%%%%%%%%%%%%%%%%%%%%%%

We begin by recalling the process of weak measurement for obtaining the weak value.  
Let $\mathcal{H}, \, \mathcal{K}$ be the Hilbert spaces associated with the system and that of the measuring device, respectively.    We wish to find the value of an observable $A$ of the system represented by a self-adjoint operator on $\mathcal{H}$.  This is done through the measurement of  
observables $Q$, $P$ of the meter device, which are represented by 
self-adjoint operators on $\mathcal{K} $ satisfying the canonical commutation relation $[Q, P] = i\hbar$ (we put $\hbar = 1$ hereafter for brevity). 
Our measurement is assumed to be of von Neumann type, in which the evolution of the composite system $\mathcal{H} \otimes \mathcal{K}$ is described by the unitary operator $e^{-igA \otimes P}$ with a coupling parameter $g \in [0, \infty)$.

Prior to the interaction, the measured system shall be prepared in some state $|\phi_{i}\rangle \in \mathcal{H}$.   Along with this \emph{preselection}, the measuring device is also prepared in a state $|\psi_{i}\rangle \in \mathcal{K}$.  The state of the composite system evolves after the interaction into $e^{-igA \otimes P} |\phi_{i}\rangle |\psi_{i}\rangle$.
We then choose a state $|\phi_{f}\rangle \in \mathcal{H}$ on which a projective measurement in $\mathcal{H}$ is performed.  Those that result in $|\phi_{f}\rangle$ shall be kept, otherwise discarded.  After this \emph{postselection}, the composite state will be disentangled into $|\phi_{f}\rangle |\psi_{f}\rangle$, where $|\psi_{f}\rangle = \langle\phi_{f}| e^{-igA \otimes P} |\phi_{i}\rangle |\psi_{i}\rangle$.

We intend to extract information of the triplet $(A, |\phi_{i}\rangle, |\phi_{f}\rangle)$ from the above measurement, and to this end we choose an observable $X = Q$ or $P$ of the measuring device and examine its shift in the expectation value  
$E_{X}(\psi) := \langle\psi |X|\psi\rangle/ \|\psi\|^{2}$ 
between the two selections:  
$
\Delta^{w}_{X}(g) := E_{X}(\psi_{f}) - E_{X}(\psi_{i})
$.
Imposing certain conditions on $|\phi_{i}\rangle$ and $|\psi_{i}\rangle$, 
both the functions $g \mapsto \Delta^{w}_{X}(g)$ are proven to be differentiable and well-defined over an open subset of $[0, \infty)$.
In particular, the functions $\Delta^{w}_{X}(g)$ are defined at $g=0$ if and only if $\langle\phi_{f} | \phi_{i}\rangle \ne 0$, in which case the derivatives at $g=0$ read
{\small
\begin{align}
&\frac{d}{dg} \Delta^{w}_{Q}(0) = \mathrm{Re}A_{w} \nonumber \\
&\qquad\quad + \left( E_{\{Q, P\}}(\psi_{i}) - 2E_{Q}(\psi_{i})E_{P}(\psi_{i}) \right) \cdot  \mathrm{Im}A_{w} , \label{eq:wv_Q} 
\\
&\frac{d}{dg} \Delta^{w}_{P}(0) = 2 \mathrm{Var}_{P}(\psi_{i}) \cdot \mathrm{Im}A_{w} ,
\label{eq:wv_P} 
\end{align}
}%
where $\{Q, P\} := QP + PQ$, $\mathrm{Var}_{X}(\psi_{i}) := E_{X^{2}}(\psi_{i}) - \left(E_{X}(\psi_{i})\right)^{2}$ the variance of $X$ on $|\psi_{i}\rangle$ and 
{\small
\begin{align}
A_{w}  := \frac{\langle \phi_{f} | A | \phi_{i} \rangle}{\langle \phi_{f} | \phi_{i} \rangle},
\end{align}
}%
a complex valued quantity called the \emph{weak value}.

From the weak measurement described above, the real and imaginary part of $A_{w}$ are obtained, in theory, with arbitrary accuracy by letting $g \to 0$.   The fact that $A_{w} $ is obtained from the rate of the shifts $\Delta^{w}_{X}(g)$ at $g=0$ justifies the term \lq weak value\rq. 
Incidentally, by imposing stricter conditions on the preselected state $|\phi_{i}\rangle$, higher order differentiability of the shifts can also be ensured, and this leads to the notion of \lq higher order weak values\rq, whose formulae are obtained analogously.

Before discussing the complications in actual measurement processes, we note that the weak value $A_{w} $ can take any arbitrary complex value by an appropriate choice of states in the two selections.
Indeed, observing that $A|\phi_{i}\rangle = E_{A}(\phi_{i})|\phi_{i}\rangle +\sqrt{\mathrm{Var}_{A}(\phi_{i})}
|\chi\rangle$ holds for any normalized $|\phi_{i}\rangle$ with $|\chi\rangle$ being a normalized vector orthogonal to $|\phi_{i}\rangle$, one can choose the postselection as $|\phi_{f}\rangle = c\, |\phi_{i}\rangle + |\chi\rangle$ with $0 \neq c \in \mathbb{C}$ to find
{\small
\begin{equation}
\frac{\langle \phi_{f} | A | \phi_{i} \rangle}{\langle \phi_{f} | \phi_{i} \rangle}
	= E_{A}(\phi_{i}) + \frac{\sqrt{\mathrm{Var}_{A}(\phi_{i})}}{c^*}.
\end{equation}
}%
Clearly, one can change freely the value of $A_{w} $ by choosing $c$ in $|\phi_{f}\rangle$ appropriately, unless 
$|\phi_{i}\rangle$ happens to be an eigenstate of $A$ for which $\mathrm{Var}_{A}(\phi_{i}) = 0$.
This is a remarkable property of the weak value and can be contrasted to the conventional eigenvalue and expectation value which are always real-valued, and bounded when $A$ is bounded.

%%%%%%%%%%%%%%%%%%%%%%%%%%%%%%
\phantom{}
\paragraph*{Conventional Measurement and Uncertainty.}
%%%%%%%%%%%%%%%%%%%%%%%%%%%%%%

In order to analyze the merit of weak measurement, we first recall the conventional (indirect) projective measurement, which is obtained by omitting the postselection in the weak measurement process. In this case, defining the shift by
$\Delta^c_X(g) := E_{\mathrm{Id} \otimes X}(e^{-igA \otimes P} \phi_{i} \otimes \psi_{i}) - E_{X}(\psi_{i})$,
one verifies
{\small
\begin{equation}\label{eq:cm}
\Delta^c_Q(g) = g \cdot E_{A}(\phi_{i}),
\end{equation}
}%
while $\Delta^c_P(g) = 0$.  Interestingly,  
for any orthonormal basis $\mathcal{B}$ of the system $\mathcal{H}$, one finds
{\small
\begin{gather}
\sum_{|\phi_{f}\rangle \in \mathcal{B}} r(\phi_{i}\to\phi_{f}) \cdot \Delta^{w}_{X}(g) = \Delta^c_X(g),
\label{eq:wqav} 
\end{gather}
}%
where
{\small
\begin{equation}\label{eq:sr}
r(\phi_{i}\to\phi_{f}) := \frac{ \left\|\left( \ket{\phi_{f}}\!\!\bra{\phi_{f}} \otimes \mathrm{Id} \right) e^{-igA \otimes P} \phi_{i}\otimes\psi_{i} \right\|^{2}}{\left\|e^{-igA \otimes P} \phi_{i}\otimes\psi_{i} \right\|^{2}}
\end{equation}
}%
is the survival rate of the postselection process in weak measurement, which tends to $\left|\langle\phi_{f} | \phi_{i}\rangle\right|^{2}$ for $g \to 0$. 
The relation \eqref{eq:wqav}  shows that, 
even for nonvanishing $g$, the shift of the weak measurement  $\Delta^{w}_{X}(g)$ reduces to the shift of the conventional measurement  $\Delta^c_{X}(g)$, that is, 
the effect of postselections disappears completely, when it is averaged over with their corresponding survival rates.

However, the aforementoined idealized measurement processes are not quite possible to implement in practice, due to 
technical/intrinsic constraints.   For instance, in measuring $X$ under the given state $|\psi\rangle$ one has the {\it systematic uncertainty} $\delta_{X} \geq 0$ arising from various sources including the finite resolution of the measuring device and its imperfect calibration.   
Besides, one has the {\it statistical uncertainty} arising from the finiteness of the number $N$ of repeated measurements 
actually performed.
To treat these uncertainties explicitly, 
let $\{ \tilde{X}_{1}, \dots, \tilde{X}_{N} \}$ be the outcomes obtained by the measurements of $X$.  
The notion of systematic uncertainty implies that there exists a sequence of \lq ideal outcomes\rq\ $X_{n} \in [\tilde{X}_{n} - \delta_{X}, \tilde{X}_{n} + \delta_{X}]$ for 
$n = 1, \ldots, N$ whose average approaches the expectation value for large $N$, that is, the error 
$\kappa^{N}_X(\psi) := \left| \sum_{n=1}^{N} {X}_{n} /N - E_{X}(\psi)\right|$
almost certainly vanishes as $\lim_{N \to \infty} \kappa^{N}_X(\psi) = 0$ (Law of Large Numbers).
For the outcomes $\tilde{X}_{n}$ with finite $N$, the triangle inequality yields
{\small
\begin{equation}\label{ineq:uncert}
\left| \frac{\sum_{n=1}^{N} \tilde{X}_{n}}{N} - E_{X}(\psi)\right| \leq \delta_{X} + \kappa^{N}_X(\psi).
\end{equation}
}%
Note that $\delta_{X}$ is intrinsic to the measurement setup and is independent of the state $|\psi\rangle$ while
$\kappa^{N}_X(\psi)$ is dependent on the statistical ensemble represented by $|\psi\rangle$.

Since the measurement outcomes $\tilde{X}_{n}$ are intrinsically probabilistic, by invoking 
Chebyshev's inequality \cite{Beasley_2004} we learn that the probability of obtaining 
the error $\kappa^{N}_X(\psi)$ to be less than a value $\kappa$ is bounded as
{\small
\begin{equation}
{\rm Prob}\left[\kappa^{N}_X(\psi) \le \kappa\right] \ge 1 - \frac{\mathrm{Var}_{X}(\psi)}{N\kappa^{2}}.
\label{prounc}
\end{equation}
}%
If one demands that the lower bound (the r.h.s.~of \eqref{prounc}) be a desired value 
$\eta$, then by solving $\kappa$ in favor of $\eta$, one can rewrite the r.h.s.~of  \eqref{ineq:uncert}  as
$
\epsilon_{X}(\eta) := \delta_{X} + \sqrt{\mathrm{Var}_{X}(\psi)/N(1-\eta)}
$
which
is specified by the probability $\eta$. 
This gives the inequality  \eqref{ineq:uncert} the meaning that   
the deviation of the value estimated from the measured outcomes $\{ \tilde{X}_{1}, \dots, \tilde{X}_{N} \}$ from $E_{X}(\psi)$
is guaranteed to be less than $\epsilon_{X}(\eta)$ 
with probability greater than $\eta$.  

Now, for the conventional measurement, suppose that $N_0$ identical sets of the composite system $|\phi_{i}\rangle|\psi_{i}\rangle$ are prepared by preselection. We collect the outcomes $\tilde{Q}^c_{n}$  of measurements of $Q$ for the meter {\it after} the interaction, and thereby obtain
$\tilde{\Delta}^c_{Q}(g) := \sum_{n=1}^{N_{0}} \tilde{Q}^c_{n}/N_{0} - E_{Q}(\psi_{i})$.
Equation \eqref{eq:cm} implies  that the ratio
$
{\tilde{\Delta}^c_{Q}(g)}/{g}
$
can be regarded as the value of $E_{A}(\phi_{i})$ estimated from the experiment, and since
$\mathrm{Var}_{\mathrm{Id} \otimes Q}(e^{-igA \otimes P} \phi_{i} \otimes \psi_{i}) = \mathrm{Var}_{Q}(\psi_i) + g^{2} \mathrm{Var}_{A}(\phi_i)$,
the accuracy $\vert {\tilde{\Delta}^c_{Q}(g)}/{g} - E_{A}(\phi_{i}) \vert$ of the estimation from the ratio is evaluated by the uncertainty,
{\small
\begin{equation}\label{eq:uncert_CM}
\epsilon^c_Q(\eta ; g, \psi_{i}) := \frac{\delta_Q}{g} + \sqrt{\frac{\mathrm{Var}_{Q}(\psi_i) + g^{2} \mathrm{Var}_{A}(\phi_i)}{g^{2}N_{0}(1-\eta)}}.
\end{equation}
}%
Observe that, while the uncertainty $\epsilon^c_Q(\eta ; g, \psi_{i})$ in the conventional measurement is in general dependent on the initial state $|\psi_{i}\rangle$ of the meter,  the dependence is washed away in the strong coupling limit
$g \to \infty$ where 
the uncertainty tends to the statistical uncertainty $\sqrt{\mathrm{Var}_{A}(\phi_i) / N_{0}(1-\eta) }$ of the system alone.

%%%%%%%%%%%%%%%%%%%%%%%%%%%%%%
\phantom{}
\paragraph*{Weak Measurement and Uncertainty.}
%%%%%%%%%%%%%%%%%%%%%%%%%%%%%%

Turning to the weak measurement, suppose that $N$ out of $N_0$ identically prepared sets of the composite system $|\phi_{i}\rangle|\psi_{i}\rangle$ survived the postselection process.
After the postselection we collect all the outcomes $\tilde{X}^w_{n}$ measured for the final state $|\psi_{f}\rangle$ of the meter and thereby obtain
$\tilde{\Delta}^{w}_{X}(g) := \sum_{n=1}^{N} \tilde{X}^w_{n}/N - E_{X}(\psi_{i})$.
Specializing to the case $X = Q$ with the initial state $|\psi_{i}\rangle$ of the meter satisfying $E_{\{Q, P\}}(\psi_{i}) - 2E_{Q}(\psi_{i})E_{P}(\psi_{i}) = 0$ for simplicity,
the relation \eqref{eq:wv_Q} and Taylor's theorem imply that the ratio
$
{\tilde{\Delta}^{w}_{Q}(g)}/{g}
$
in the limit $g \to 0$ can be regarded as the estimated value of $\mathrm{Re}A_{w} $ 
from the experiment.   In the actual experiment, however, the coupling constant $g$ should be kept nonvanishing, and hence we have
{\footnotesize
\begin{align}
\left| \frac{\tilde{\Delta}^{w}_{Q}(g)}{g} - \mathrm{Re}A_{w}  \right|
	&= \left| \frac{\tilde{\Delta}^{w}_{Q}(g) - \Delta^{w}_{Q}(g)}{g} 
 + \left( \frac{\Delta^{w}_{Q}(g)}{g} - \mathrm{Re}A_{w}  \right)\right| \nonumber \\
	&\leq \frac{\delta_{Q}}{g} + \frac{\kappa^{N}_Q(\psi_f)}{g} 
	 + \left| \frac{\Delta^{w}_{Q}(g)}{g} - \mathrm{Re}A_{w} \right|,
\label{ineqp}
\end{align}
}%
in place of \eqref{ineq:uncert}.  In addition, 
since the process of obtaining $N$ out of $N_{0}$ outcomes generally depends on the postselection in relation to the preselection, we must also take the survival rate \eqref{eq:sr} into account. 
To discuss the uncertainties along this more realistic line, note that the probability of $N$ survived out of $N_0$ is given by
the binomial distribution $\mathrm{Bi}\left[N; N_0,r\right] :=  \binom{N_0}{N} r^{N} \left( 1 - r \right)^{N_0-N}$ with $r$ given by the survival rate \eqref{eq:sr}.
To each of these $N$ outcomes, inequality \eqref{ineqp}
holds with the lower bound of the probability \eqref{prounc}.   Thus, the average probability that the measurement 
yields outcomes within the statistical error $\kappa$ is given by the sum over all possible $N$,
{\footnotesize
\begin{align}
\label{avpro}
{\Pi}^{N_0}_{Q}(\kappa) 
:= \sum_{N=1}^{N_0} \mathrm{Bi}\left[N; N_0, r(\phi_{i}\to\phi_{f})\right]  \max \left[ \left( 1-\frac{\mathrm{Var}_{Q}(\psi_{f})}{N\kappa^{2}} \right), 0 \right].
\end{align}
}
To ensure the overall uncertainty level by some $\eta > 0$, we may put $\eta = {\Pi}^{N_0}_{Q}(\kappa)$.  This relation can be solved for $\kappa$ to obtain the inverse $\kappa_Q^{N_0}(\eta) := [\Pi^{N_0}_{Q}]^{-1}(\eta)$, since each term in the sum \eqref{avpro} is a continuous and monotonically increasing function in $\kappa$.  
From this, the uncertainty of estimating $\mathrm{Re}A_{w}$ by weak measurement is given by
{\small
\begin{align}\label{eq:uncert_WM}
\epsilon^{w}_{Q}(\eta; g, \psi_{i}) := \frac{\delta_{Q}}{g} + 
\frac{\kappa^{N_{0}}_Q(\eta)}{g} + \left| \frac{\Delta^{w}_{Q}(g)}{g} - \mathrm{Re}A_{w}\right|.
\end{align}
}%
The third term in \eqref{eq:uncert_WM}, which is absent for the conventional measurement \eqref{eq:uncert_CM}, is due to the nonlinearity of the shift with respect to $g$, which cannot be ignored for nonvanishing $g$ in realistic settings. An analogous argument holds for the estimation of $\mathrm{Im}A_{w} $ with the choice $X = P$.

We thus have obtained a framework for handling both the ideal and realistic measurement in terms of uncertainty, where the ideal case arises  in the limit $\delta_{X}\to 0$, $N_{0}\to \infty$ (and $g \to 0$ for weak measurement) for which the uncertainties \eqref{eq:uncert_CM} and \eqref{eq:uncert_WM} vanish for all $\eta$.
In passing, we mention that the uncertainties are invariant under translation along the real axis  $A \to A + t$ for $t \in \mathbb{R}$.
As for scaling $A \to rA,\  g \to g/r$ for $r > 0$,
we just have $\epsilon^{c,w}_{X}(\eta; g, \psi_{i}) \to r\epsilon^{c,w}_{X}(\eta; g, \psi_{i})$ 
as expected.

%%%%%%%%%%%%%%%%%%%%%%%%%%%%%%
\phantom{}
\paragraph*{Merit of Weak Measurement and Amplification.}
%%%%%%%%%%%%%%%%%%%%%%%%%%%%%%

Now we address the question of whether the weak measurement has an advantage over the conventional one for obtaining, {\it e.g.}, the expectation value of $A$.
Obviously, since the weak measurement requires postselection, one cannot fully exploit all samples prepared prior to the measurement.  This yields a larger statistical uncertainty
which could quickly become uncontrollably large for higher $\eta$. 
Even in the ideal limit $N_{0} \to \infty$ where the statistical uncertainty vanishes, there still remains nonlinearity, which is nonexistent in the conventional case.   By comparing $\eqref{eq:uncert_CM}$ and $\eqref{eq:uncert_WM}$, we learn
that, as long as the uncertainty is concerned, there is no technical merit for adopting the weak measurement. 
However, the true merit of weak measurement can arise in the situation where the amplification of the weak value outside of the numerical range $W(A) := \{\langle \phi |A| \phi\rangle : \|\phi\| = 1 \}$ is available.

To see this by a simple example,
consider a situation where the strength $g$ of interaction cannot be made large enough to curb the systematic error.  For instance, if $A$ is bounded and the numerical range $W(A)$ is confined in a much smaller region than $\delta/g$ for available $g$, then the estimated value of any $E_{A}(\phi) \in W(A)$ is completely obscured by $\delta/g$, in which case the projective measurement reveals no meaningful information of the system.   In contrast, in the weak measurement the range of the weak value spans the whole complex plane, and one can arrange the selections such that  $\mathrm{Re}A_{w}$ is amplified outside of $W(A)$
rendering the systematic error negligible compared to $\mathrm{Re}A_{w}$.
To be more specific, 
suppose that one probes the very existence of a physical effect by looking at the shift of the meter in the measurement, as in the case of the experiments \cite{Hosten_2008, Dixon_2009}. 
We can conclude that the effect exists with confidence $\eta$ when
{\small
\begin{equation}
\label{eq:uncert_PMSC}
\epsilon^c_Q(\eta ; g, \psi_{i}) \leq \vert E_{A}(\phi_{i})\vert,
\end{equation}
}%
{\small
\begin{equation}
\label{eq:uncert_WMSC}
\epsilon^{w}_Q(\eta ; g, \psi_{i}) \leq \vert \mathrm{Re}A_{w}\vert , \quad \epsilon^{w}_P(\eta ; g, \psi_{i}) \leq \vert \mathrm{Im}A_{w}\vert ,
\end{equation}
}%
hold for respective measuremens, which are all sufficient conditions for distinguishing the coupling $g$ from $g=0$.  In both of these cases, we may call the measurements \emph{significant with confidence $\eta$}. Clearly, for precision measurement the weak measurement becomes superior when \eqref{eq:uncert_PMSC} is broken while \eqref{eq:uncert_WMSC} can be maintained 
through the amplification of $A_{w}$.

%%%%%%%%%%%%%%%%%%%%%%%%%%%%%%
\phantom{}
\paragraph*{Gaussian Model with Two-Point Spectrum.}
%%%%%%%%%%%%%%%%%%%%%%%%%%%%%%

For illustration, we consider the model in which $A$ has a discrete spectrum consisting of two distinct values $\{\lambda_{1}, \lambda_{2}\}$ and the initial state
$\psi_{i}$ of the meter $\mathcal{K} = L^{2}(\mathbb{R})$ is given by normalized Gaussian wave functions $\psi_{i}(x) = ( 1/\pi d^{2})^{\frac{1}{4}} \exp(-\frac{x^{2}}{2d^{2}})$ centered at $x = 0$ with width $d > 0$.
Despite its simplicity, this model is sufficiently general in the sense that it covers most of the recent experiments of weak measurement \cite{Hosten_2008, Dixon_2009} as well as the recent work \cite{Wu_2011, Koike_2011, Nakamura_2012} where a full order calculation of the shift for $A$ satisfying $A^{2} = \mathrm{Id}$ is performed, which is equivalent to $\{\lambda_{1}, \lambda_{2}\} = \{ -1, 1\}$ under our setting.
Identifying the usual position and momentum operators on $L^{2}(\mathbb{R})$ with $Q=\Hat{x}$ and $P=\Hat{p}$, and using
the shorthand, $\Lambda_{m} := (\lambda_{1} + \lambda_{2})/{2}$ and $\Lambda_{r} := (\lambda_{2} - \lambda_{1})/{2}$, we find 
{\small
\begin{align}
	\Delta^{w}_{Q}(g) &= g \cdot \frac{\mathrm{Re} A_{r} }{1 + a\left( 1 - e^{-g^{2}\Lambda_{r}^{2}/d^{2}} \right)} + g \cdot \Lambda_{m}, \label{eq:Gauss_Shift_Q} \\
	\Delta^{w}_{P}(g) &= \frac{g}{d^{2}} \cdot \frac{ \mathrm{Im} A_{r} e^{-g^{2}\Lambda_{r}^{2}/d^{2}}}{1 + a \left( 1 - e^{-g^{2}\Lambda_{r}^{2}/d^{2}} \right)} \label{eq:Gauss_Shift_P},
\end{align}
}%
where 
$
A_{r} := A_{w} - \Lambda_{m}
$,
and
$a := \frac{1}{2} \left(\vert A_{r}\vert^{2}/\Lambda_{r}^{2} - 1 \right)$
is a parameter corresponding to the amplification rate.
One then verifies that the above functions are of class $C^{\infty}$ with respect to $g$, and the estimated values ${\Delta^{w}_{Q}(g)}/{g}$ and ${\Delta^{w}_{P}(g)}/{(g/d^{2})}$
tend to $\mathrm{Re} A_{w}$ and $\mathrm{Im} A_{w}$, respectively, in the weak limit $g \to 0$.

The uncertainties for this model can be analytically obtained based on \eqref{eq:uncert_WM} and its counterpart for measuring $\mathrm{Im}A_{w}$ (see Supplemental Material).
At this point, observe that both the statistical and the nonlinear terms in the overall uncertainties are dependent on $g$ and $d$ only through the combination $g/d$.   Thus, instead of considering the weak limit $g \to 0$, one may equally consider the broad limit of the width $d \to \infty$ to obtain the weak value $A_{w}$ from ${\Delta^{w}_{Q}(g)}/{g}$ and ${\Delta^{w}_{P}(g)}/{(g/d^{2})}$.  In other words, the aim of weak measurement to obtain the weak value can be achieved with a coupling $g$ which is not weak at all.
We also remark that any choice of the spectrum $\{\lambda_{1}, \lambda_{2}\}$ can change into one another by some combination of translation and scaling. 
In fact, on account of the aforementioned properties of the uncertainties under these transformations, one sees that all these models actually reduce to the simplest case $\{\lambda_{1}, \lambda_{2}\} = \{ -1, 1\}$.

\begin{figure}
\vspace{20pt}
\centering
\includegraphics[width=65mm]{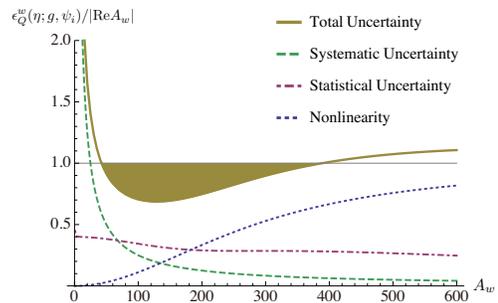}
\caption{
Ratio of uncertainty $\epsilon^{w}_{Q}(\eta; g, d)$ to the real part of 
the weak value of the spin $S_{z} = \sigma_z/2$. By amplifying the weak value out of 
its numerical range $[-1/2, 1/2]$ to $(S_{z})_{w} \approx 100$, 
the significance condition \eqref{eq:uncert_WMSC} is attained with confidence $\eta=0.95$. 
(Parameters: $\delta_{Q} = 1/2$, $N_{0}= 10^{7}$, 
$g=1/50$ and $d = 4$.)
}
\label{fig:Amplification}
\end{figure}

We now return to the problem of fulfillment of the significance condition \eqref{eq:uncert_WMSC}.
With a proper choice of measurement setups, this condition can indeed be fulfilled (while  \eqref{eq:uncert_PMSC} is broken) as shown in Fig.~\ref{fig:Amplification}.
%%%%%%%%
%%%%%%%%
Note, however, that the advantage of amplification does not come free, because the amplification requires generically a small transition amplitude $\langle\phi_{f} | \phi_{i}\rangle$, which necessitate 
a much larger number of prepared samples to suppress the statistical uncertainty compared to the conventional one. 
More importantly, the amplification enhances  also the uncertainty coming from the nonlinearity and, together with the statistical uncertainty, eventually ruins the significance condition \eqref{eq:uncert_WMSC}.   In fact, it can be proved that, for an observable $A$ with finite eigenvalues, the shifts $\Delta^{w}_{X}(g)$ for fixed $g$ and $|\psi_{i}\rangle$ are also bounded with respect to a set of pre and postselections of the system and, as a result, the ratio between the amplified weak value and the nonlinear term becomes unity as $|\mathrm{Re}A_{w}|,\ |\mathrm{Im}A_{w}| \to \infty$,
suggesting that the qualitative behaviour seen in this specific numerical demonstration is actually universal. 
Given these trade-offs, 
for probing a physical effect such as gravitational waves by means of weak measurement,  
it is vital for us to find a possible range of amplification fulfilling \eqref{eq:uncert_WMSC} where the measurement is significant.

\begin{acknowledgments}
We thank Prof. A. Hosoya for helpful discussions.
This work was supported by JSPS Grant-in-Aid for Scientific Research (C), No.~25400423. 
\end{acknowledgments}

\end{document}

%%%%%%%%    file name   %%%%%%%%%%%%%%%%%%%%
\title{Supplemental Material for ``Uncertainty of Weak Measurement and Merit of Amplification"}

\author{Jaeha Lee}
\email[Email:]{jlee@post.kek.jp}
\affiliation{Department of Physics, University of Tokyo, 7-3-1 Hongo, Bunkyo-ku, Tokyo 113-0033, Japan}

\author{Izumi Tsutsui}
\email[Email:]{izumi.tsutsui@kek.jp}
\affiliation{Theory Center, Institute of Particle and Nuclear Studies,
High Energy Accelerator Research Organization (KEK), 1-1 Oho, Tsukuba, Ibaraki 305-0801, Japan}

\date{\today}

\maketitle

This supplemental material provides proofs for the general statements made in the Letter, together with the full computational procedure for obtaining the uncertainties in the case of the Gaussian model.
To avoid unnecessary complication, the following mathematical arguments are given in a rather laxer way, although its full elaboration is quite straightforward.
Throughout this material, we denote by $\mathcal{H}$ and $\mathcal{K}$ the Hilbert spaces representing the states of the system and that of the meter, respectively. 
For the physical observable, we consider a self-adjoint operator $A$ acting on the system $\mathcal{H}$, and for the pair of observables of the meter device required in weak measurement, we use $Q$ and $P$ which are self-adjoint operators acting on $\mathcal{K}$ satisfying the canonical commutation relation $[Q, P] = i\hbar$ (we put $\hbar = 1$ hereafter for brevity). Apart from the Gaussian state analysis used for demonstration, our argument is made in a completely general setting (with the only exception that the interaction between the two quantum systems is assumed to be of von Neumann type, although this condition too can be generalized). The order of presentation, as well as notations and definitions in this supplemental material, proceed basically with that of the Letter, so that the readers can read the two in parallel.

%%%%%%%%%%%%%%%%%%%%%%%%%%%%%%%%%%%%%%%%%%%%
%%%%%%%%%%%%%%%%%%%%%%
\section{(Ideal) Weak and Conventional Measurement}
%%%%%%%%%%%%%%%%%%%%%%
%%%%%%%%%%%%%%%%%%%%%%%%%%%%%%%%%%%%%%%%%%%%

In the first section we furnish a mathematical basis for both conventional and weak measurement in the ideal setting where no error is present.   We consider a von Neumann type interaction between the two Hilbert spaces, where the composite system $\mathcal{H} \otimes \mathcal{K}$ evolves according to the unitary operator $e^{-igA \otimes P}$ with a coupling parameter $g \in [0, \infty)$. 

%%%%%%%%%%%%%%%%%%%%%%
\subsection{Weak Value and Weak Measurement}
%%%%%%%%%%%%%%%%%%%%%%

We begin by providing a rough sketch of the proof of (high order) differentiability of the shifts $\Delta^{w}_{X}(g,\psi_{i})$ of the weak measurement 
with respect to the coupling constant $g$ (formulae (1) and (2) in the Letter).
Define the collection $\{W_{g}\}_{g \in [0, \infty)}$ of transition maps of the meter by
$W_{g} : |\psi_{i}\rangle \mapsto |\psi_{f}\rangle$,
where $|\psi_{f}\rangle = \langle\phi_{f}| e^{-igA \otimes P} |\phi_{i}\rangle |\psi_{i}\rangle$ is the final state of the meter after the postselection process with normalized $|\phi_{i}\rangle, |\phi_{f}\rangle \in \mathcal{H}$. One sees that each transition map is a bounded linear operator acting on the meter $\mathcal{K}$.
Now, let
$S$ be any invariant subspace of $Q$ and $P$, {\it i.e.}, a subspace of $\mathcal{K}$ satisfying $S \subset \mathrm{dom}(Q) \cap  \mathrm{dom}(P)$ and  $QS \subset S$, $PS \subset S$, where $\mathrm{dom}(X)$ denotes the domain of the operator $X$. In such settings, by choosing $|\phi_{i}\rangle \in \mathrm{dom}(A^{n})$ ($n \in \mathbb{N}$), one verifies that the function $g \mapsto |\psi_{f}\rangle = W_{g}|\psi_{i}\rangle$ is of class $C^{n}$, hence in particular
\begin{align}
\left.\frac{d^{n}}{dg^{n}} |\psi_{f}\rangle\right|_{g=0} &= \left.\frac{d^{n}}{dg^{n}}W_{g} |\psi_{i}\rangle\right|_{g=0} \nonumber \\
    &= (-i)^{n} \langle\phi_{f}| A^{n} |\phi_{i}\rangle \cdot P^{n} |\psi_{i}\rangle
\end{align}
holds at $g=0$. Moreover, it is well-known that for any operator $X$ the map
$\psi \mapsto E_{X}(\psi) :=\langle\psi| X |\psi\rangle / \|\psi\|^{2}$
is $n$ times Fréchet-Differentiable on $\mathrm{dom}(X^{n})\setminus \{0\}$ ($n \in \mathbb{N}$). In particular for $n=1$, one has
{\small
\begin{align}
DE_{X}(\psi)\{\upsilon\} &:= \lim_{h \to 0} \frac{E_{X}(\psi + h\upsilon) - E_{X}(\psi)}{h} \nonumber \\
    &= 2\mathrm{Re} \left[ \frac{\left\langle \psi |  X - E_{X}(\psi)  | \upsilon \right\rangle}{\|\psi\|^{2}} \right].
\end{align}
}%
Combining both of these facts, by choosing $|\psi_{i}\rangle \in S$ and $|\phi_{i}\rangle \in \mathrm{dom}(A^{n})$, we see that the shift of the meter for both $X=Q,P$,
\begin{align}
\Delta^{w}_{X}(g,\psi_{i}) &:= E_{X}(\psi_{f}) - E_{X}(\psi_{i}) \nonumber \\
    &= E_{(|\phi_{f}\rangle\langle\phi_{f}| \otimes X)}(e^{-igA \otimes P} \phi_{i}\otimes\psi_{i}) - E_{X}(\psi_{i}).
\end{align}
is well-defined for $g$ contained in some open subset of $[0, \infty)$, and is of class $C^{n}$. Observing that $W_{g=0} = \langle\phi_{f} | \phi_{f}\rangle \cdot \mathrm{Id}$, if we moreover require the condition $\langle\phi_{f} | \phi_{i}\rangle \neq 0$, the above shift function is defined at $g=0$, in which case the measurement setup is called \emph{non-orthogonal weak measurement}. In this case, applying the chain rule, the differential coefficients at $g=0$ read
{\small
\begin{align}
\left.\frac{d}{dg}E_{X}(\psi_{f})\right|_{g=0} &= DE_{X}(W_{0}\psi_{i}) \circ \left.\frac{d}{dg} |\psi_{f}\rangle\right|_{g=0} \nonumber \\
    &= 2\mathrm{Im} \left[ \frac{\left\langle \langle \phi_{f} | \phi_{i}\rangle \cdot \psi_{i} |\left( X - E_{X}(\psi_{i}) \right)| \langle\phi_{f}| A |\phi_{i}\rangle \cdot P \psi_{i} \right\rangle}{|\langle \phi_{f} | \phi_{i}\rangle|^{2} \cdot \|\psi_{i}\|^{2}} \right] \nonumber \\
    &= 2\mathrm{Im} \left[  A_{w} \cdot \frac{\left\langle \psi_{i} | X - E_{X}(\psi_{i}) | P \psi_{i} \right\rangle}{\|\psi_{i}\|^{2}} \right] \nonumber \\
    &= 2\mathrm{Im} \left[  A_{w} \cdot \left\{ \frac{\left\langle \psi_{i} |X P| \psi_{i} \right\rangle}{\|\psi_{i}\|^{2}} - E_{X}(\psi_{i})E_{P}(\psi_{i}) \right\} \right] \nonumber \\
    &=  2\mathrm{Im} \left[ \frac{\left\langle  \psi_{i} | X P | \psi_{i} \right\rangle}{\|\psi_{i}\|^{2}} \right] \cdot \mathrm{Re}A_{w} + \left\{ 2\mathrm{Re} \left[ \frac{\left\langle \psi_{i} |X P| \psi_{i} \right\rangle}{\|\psi_{i}\|^{2}} \right] - 2E_{X}(\psi_{i})E_{P}(\psi_{i})  \right\} \cdot \mathrm{Im}A_{w},
\end{align}
}%
for $n=1$, where
\begin{align}
\label{eq:wvalue}
A_{w} := \frac{\langle\phi_{f}| A |\phi_{i}\rangle}{\langle\phi_{f} | \phi_{i}\rangle}.
\end{align}
is the weak value of the observable $A$.
Hence, we learn that
{\small
\begin{align}
\frac{d}{dg}\Delta^{w}_{Q}(0,\psi_{i}) &= \mathrm{Re}A_{w} + \left( E_{\{Q,P\}}(\psi_{i}) - 2E_{Q}(\psi_{i})E_{P}(\psi_{i})  \right) \cdot \mathrm{Im}A_{w}, \label{eq:wv_Q}  \\
\frac{d}{dg}\Delta^{w}_{P}(0,\psi_{i}) &= 2\mathrm{Var}_{P}(\psi_{i}) \cdot \mathrm{Im}A_{w},\label{eq:wv_P} 
\end{align}
}%
where we have used $\left\langle \psi|Q P|\psi \right\rangle = \left\langle \psi|P Q|\psi \right\rangle + i\| \psi \|^{2}$, {\it i.e.}, $2\, \mathrm{Im} \left[ \left\langle \psi_{i}|QP| \psi_{i} \right\rangle / \|\psi_{i}\|^{2} \right] = 1$.
`Higher order weak values' ($n \geq 2$) can be analogously obtained.

%%%%%%%%%%%%%%%%%%%%%%
\subsection{Conventional (Indirect Projective) Measurement}
%%%%%%%%%%%%%%%%%%%%%%

Before discussing the conventional (indirect projective) measurement, we first note the following lemma, which proves itself to be useful in later arguments.

\begin{Lem}\label{eq:com}
For any $g \in \mathbb{R}$, the equation
{\small
\begin{equation}
(\mathrm{Id} \otimes Q) \circ e^{-ig(A \otimes P)} = e^{-ig(A \otimes P)} \circ \left( (\mathrm{Id} \otimes Q) + g (A \otimes \mathrm{Id} ) \right)
\end{equation}
}%
holds.
\end{Lem}
\begin{Prf}
A heuristic argument is given. It suffices to see that the equation
{\small
\begin{equation}
(\mathrm{Id} \otimes Q) \circ (A \otimes P)^{n} = (A \otimes P)^{n} \circ (\mathrm{Id} \otimes Q) + in (A \otimes P)^{n-1} \circ (A \otimes \mathrm{Id})
\end{equation}
}%
holds for any $n \in \mathbb{N}$.
For $n=1$, one has
{\small
\begin{align}
(\mathrm{Id} \otimes Q) \circ (A \otimes P) &= (A \otimes QP) \nonumber \\
	&= (A \otimes (PQ + i\mathrm{Id})) \nonumber \\
	&= (A \otimes P) \circ (\mathrm{Id} \otimes Q) + i(A \otimes \mathrm{Id}).
\end{align}
}%
Suppose the equation holds for $n \geq 1$. One has
{\small
\begin{align}
(\mathrm{Id} \otimes Q) \circ (A \otimes P)^{n+1} &= (\mathrm{Id} \otimes Q) \circ (A \otimes P)^{n} \circ (A \otimes P) \nonumber \\
	&= \left\{ (A \otimes P)^{n} \circ (\mathrm{Id} \otimes Q) + in (A \otimes P)^{n-1} \circ (A \otimes \mathrm{Id}) \right\} \circ (A \otimes P) \nonumber \\
	&= (A \otimes P)^{n} \circ (\mathrm{Id} \otimes Q) \circ (A \otimes P) + in (A \otimes P)^{n} \circ (A \otimes \mathrm{Id}) \nonumber \\
	&= (A \otimes P)^{n+1} \circ (\mathrm{Id} \otimes Q) +i(A \otimes P)^{n} (A \otimes \mathrm{Id}) + in (A \otimes P)^{n} \circ (A \otimes \mathrm{Id}) \nonumber \\
	&= (A \otimes P)^{n+1} \circ (\mathrm{Id} \otimes Q) + i(n+1) (A \otimes P)^{n} \circ (A \otimes \mathrm{Id}).
\end{align}
}%
One thus has the desired equation by induction. Then,
{\small
\begin{align}
(\mathrm{Id} \otimes Q) \circ e^{-ig(A \otimes P)} &= \sum_{n=0}^{\infty} \frac{(-ig)^{n} (\mathrm{Id} \otimes Q) \circ (A \otimes P)^{n}}{n!} \nonumber \\
	&= \sum_{n=0}^{\infty} \frac{(-ig)^{n} (A \otimes P)^{n} \circ (\mathrm{Id} \otimes Q)}{n!} + \sum_{n=0}^{\infty} \frac{(-ig)^{n} in (A \otimes P)^{n-1} \circ (A \otimes \mathrm{Id})}{n!} \nonumber \\
	&= e^{-ig(A \otimes P)} \circ (\mathrm{Id} \otimes Q) + g \sum_{n=0}^{\infty} \frac{(-ig)^{n-1}  (A \otimes P)^{n-1} \circ (A \otimes \mathrm{Id})}{(n-1)!} \nonumber \\ 
	&= e^{-ig(A \otimes P)} \circ (\mathrm{Id} \otimes Q) + g e^{-ig(A \otimes P)} \circ (A \otimes \mathrm{Id}).
\end{align}
}%
\qed
\end{Prf}

As we did in the previous subsection, let $|\phi_{i}\rangle \in \mathrm{dom}(A)$, $|\psi_{i}\rangle \in S$ be preselected states of the system $\mathcal{H}$ and that of the meter $\mathcal{K}$, respectively.  Using the previous lemma we find
{\small
\begin{align}
E_{\mathrm{Id} \otimes Q}(e^{-igA \otimes P}\phi_{i} \otimes \psi_{i}) :=& \frac{\langle e^{-igA \otimes P} \phi_{i} \otimes \psi_{i} | \mathrm{Id} \otimes Q | e^{-igA \otimes P} \phi_{i} \otimes \psi_{i} \rangle}{\|e^{-igA \otimes P}\phi_{i} \otimes \psi_{i}\|^{2}} \nonumber \\
	=& \frac{\left\langle \phi_{i} \otimes \psi_{i} |  (\mathrm{Id} \otimes Q) + g (A \otimes \mathrm{Id})  | \phi_{i} \otimes \psi_{i} \right\rangle}{\|\phi_{i} \otimes \psi_{i}\|^{2}} \nonumber \\
	=& E_{Q}(\psi_{i}) + gE_{A}(\phi_{i}),
\end{align}
}%
and
{\small
\begin{align}
E_{(\mathrm{Id} \otimes Q)^{2}}(e^{-igA \otimes P}\phi_{i} \otimes \psi_{i}) :=& \frac{\langle e^{-igA \otimes P} \phi_{i} \otimes \psi_{i} | (\mathrm{Id} \otimes Q)^{2} | e^{-igA \otimes P} \phi_{i} \otimes \psi_{i} \rangle}{\|e^{-igA \otimes P}\phi_{i} \otimes \psi_{i}\|^{2}} \nonumber \\
	=& \frac{\left\langle \phi_{i} \otimes \psi_{i} | \left( (\mathrm{Id} \otimes Q) + g (A \otimes \mathrm{Id}) \right)^{2} | \phi_{i} \otimes \psi_{i} \right\rangle}{\|\phi_{i} \otimes \psi_{i}\|^{2}} \nonumber \\
	=& E_{Q^{2}}(\psi_{i}) + 2gE_{A \otimes Q}(\phi_{i} \otimes \psi_{i}) + g^{2}E_{A^{2}}(\phi_{i}) \nonumber \\
	=& E_{Q^{2}}(\psi_{i}) + 2gE_{Q}(\psi_{i})E_{A}(\phi_{i}) + g^{2}E_{A^{2}}(\phi_{i}).
\end{align}
}%
Now, defining the shifts of the meter also for the conventional measurement case by
{\small
\begin{equation}
\Delta_{X}^{c}(g,\psi_{i}) := E_{\mathrm{Id} \otimes X}(e^{-igA \otimes P}\phi_{i}\otimes\psi_{i}) - E_{X}(\psi_{i}),
\end{equation}
}%
for $X = Q, P$, we obtain
{\small
\begin{equation}
\label{eq:shift_cm}
\Delta_{Q}^{c}(g,\psi_{i}) = g \cdot E_{A}(\phi_{i}),
\end{equation}
}%
which is formula (5) in the Letter, and also $\Delta_{P}^{c}(g,\psi_{i}) = 0$.
Likewise, the variance can be rewritten as
{\small
\begin{align}\label{eq:Var_Proj}
\mathrm{Var}_{\mathrm{Id} \otimes Q}(e^{-igA \otimes P} \phi_{i} \otimes \psi_{i}) :=& E_{(\mathrm{Id} \otimes Q)^{2}}(e^{-igA \otimes P}\phi_{i} \otimes \psi_{i}) - \left[E_{(\mathrm{Id} \otimes Q)}(e^{-igA \otimes P}\phi_{i} \otimes \psi_{i})\right]^{2} \nonumber \\
	=& E_{Q^{2}}(\psi_{i}) + 2gE_{Q}(\psi_{i})E_{A}(\phi_{i}) + g^{2}E_{A^{2}}(\phi_{i}) - \left[E_{Q}(\psi_{i}) + gE_{A}(\phi_{i})\right]^{2} \nonumber \\
	=& \left( E_{Q^{2}}(\psi_{i}) - E_{Q}(\psi_{i})^{2} \right) + g^{2}\left( E_{A^{2}}(\phi_{i}) - E_{A}(\phi_{i})^{2} \right) \nonumber \\
	=& \mathrm{Var}_{Q}(\psi_{i}) + g^{2} \mathrm{Var}_{A}(\phi_{i}),
\end{align}
}%
whose result is used in obtaining Eq.(10) in the Letter.

%%%%%%%%%%%%%%%%%%%%%%
\subsection{Conventional Measurement and Weak Measurement}
%%%%%%%%%%%%%%%%%%%%%%

As we have seen above, the difference between the conventional and the weak measurement appears only in the definition of the shifts $\Delta^{c}_{X}(g,\psi_{i})$ and $\Delta^{w}_{X}(g,\psi_{i})$, in which we measure either the observable $\mathrm{Id} \otimes X$ or $|\phi_{f}\rangle\langle\phi_{f}| \otimes X$, respectively, on the composite state $e^{-igA\otimes P}|\phi_{i}\rangle|\psi_{i}\rangle$ after the interaction.   The two shifts are related through the survival rate of the postselection,
{\small
\begin{align}\label{eq:sr}
r(\phi_{i} \to \phi_{f}) &:= \frac{ \left\|\left( \ket{\phi_{f}}\!\!\bra{\phi_{f}} \otimes \mathrm{Id} \right) e^{-igA \otimes P} \phi_{i} \otimes \psi_{i} \right\|^{2}}{\left\|e^{-igA \otimes P} \phi_{i} \otimes \psi_{i} \right\|^{2}} \nonumber \\
    &= \frac{\|\psi_{f}\|^{2}}{\left\|\phi_{i} \otimes \psi_{i} \right\|^{2}}.
\end{align}
}%
Indeed, after summing up various postselections belonging to an arbitrary orthonormal basis $\mathcal{B}$ of the system $\mathcal{H}$
with their corresponding survival rates, we find
{\small
\begin{align}
\label{eq:wvaexp}
\sum_{|\phi_{f}\rangle \in \mathcal{B}} r(\phi_{i} \to \phi_{f}) \cdot \Delta^{w}_{X}(g, \psi_{i}) &:= \sum_{|\phi_{f}\rangle \in \mathcal{B}} \frac{\|\psi_{f}\|^{2}}{\left\|\phi_{i} \otimes \psi_{i} \right\|^{2}} \cdot \left\{ \frac{\langle\psi_{f}|X|\psi_{f}\rangle}{\|\psi_{f}\|^{2}} - E_{X}(\psi_{i}) \right\} \nonumber \\
    &= \sum_{|\phi_{f}\rangle \in \mathcal{B}} \frac{\langle\psi_{f}|X|\psi_{f}\rangle}{\left\|\phi_{i} \otimes \psi_{i} \right\|^{2}} - E_{X}(\psi_{i}) \nonumber \\
    &= \sum_{|\phi_{f}\rangle \in \mathcal{B}} \frac{\langle e^{-igA \otimes P} \phi_{i} \otimes \psi_{i} | \left( |\phi_{f}\rangle\langle\phi_{f}| \otimes X \right) | e^{-igA \otimes P} \phi_{i} \otimes \psi_{i}\rangle}{\left\|\phi_{i} \otimes \psi_{i} \right\|^{2}} - E_{X}(\psi_{i}) \nonumber \\
    &= \frac{\langle e^{-igA \otimes P} \phi_{i} \otimes \psi_{i} | \left( \mathrm{Id} \otimes X \right) | e^{-igA \otimes P} \phi_{i} \otimes \psi_{i}\rangle}{\left\|\phi_{i} \otimes \psi_{i} \right\|^{2}} - E_{X}(\psi_{i}) \nonumber \\
    &= E_{\mathrm{Id} \otimes X}(e^{-igA \otimes P}\phi_{i} \otimes \psi_{i}) - E_{X}(\psi_{i}) \nonumber \\
    &=: \Delta_{X}^{c}(g,\psi_{i})
\end{align}
}%
which is formula (6) in the Letter. In particular, since we have \eqref{eq:wv_Q}, \eqref{eq:wv_P}, \eqref{eq:shift_cm} and $\lim_{g \to 0} r(\phi_{i} \to \phi_{f}) = |\langle \phi_{f}|\phi_{i}\rangle|^{2}$, we obtain, as a special case, the following relation
{\small
\begin{equation}
\sum_{|\phi_{f}\rangle \in \mathcal{B}} |\langle \phi_{f}|\phi_{i}\rangle|^{2} \cdot A_{w} = E_{A}(\phi_{i}),
\end{equation}
}%
which is commonly known as a formula providing the relation between the weak value and the expectation value.   The relation \eqref{eq:wvaexp} is a little more general and shows that
the effect of postselections disappears completely after averaging over the postselections even for nonvanishing $g$.

%%%%%%%%%%%%%%%%%%%%%%%%%%%%%%%%%%%%%%%%%%%%
%%%%%%%%%%%%%%%%%%%%%%
\section{Scaling and Translation Properties of the Uncertainty}
%%%%%%%%%%%%%%%%%%%%%%
%%%%%%%%%%%%%%%%%%%%%%%%%%%%%%%%%%%%%%%%%%%%

Based on the argument given in the Letter, the formulae for the uncertainties (Eqs.(10) and (13) in the Letter) for both the conventional and the weak measurement model are obtained. 
Now, it is fairly straightforward to prove both its scaling and translation properties mentioned in the Letter. 
Firstly, it is obvious by definition that the transition maps $\{W_{g}\}_{g \in [0, \infty)}$ are invariant under scaling $A \to r \cdot A,\ g \to g/r$ for $r > 0$.  It follows that the components constituting the uncertainties $\epsilon^{w}_{X}(\eta;g,\psi_{i})$, {\it i.e.}, the survival rate $r(\phi_{i} \to \phi_{f})$, the shifts $\Delta^{w}_{X}(g, \psi_{i})$ and the variances $\mathrm{Var}_{X}(\psi_{f})$,  are all invariant under the scaling, and consequently we find from the definition of $\epsilon^{w}_{X}(\eta;g,\psi_{i})$ that $\epsilon^{w}_{X}(\eta;g,\psi_{i}) \to r \cdot \epsilon^{w}_{X}(\eta;g,\psi_{i})$ as $A \to r \cdot A,\ g \to g/r$.

As for translation $A \to A + t$ for $t \in \mathbb{R}$, we see that
\begin{align}
e^{-ig((A+t)\otimes P)} &= e^{-igt(\mathrm{Id} \otimes P)} \circ e^{-ig(A\otimes P)} \nonumber \\
	&= \left(\mathrm{Id} \otimes e^{-igt P}\right) \circ e^{-ig(A\otimes P)},
\end{align}
in which case we have $W_{g} \to e^{-igtP} \circ W_{g}$ as $A \to A + t$. 
Since $Q^{n} \circ e^{-igt P} = e^{-igt P} \circ \left( gt + Q\right)^{n}$ holds for $n \geq 0$, we have 
$E_{Q^{n}}(\psi_{f}) \to E_{(Q+gt)^{n}}(\psi_{f})$
as $A \to A + t$, and hence
{\small
\begin{align}
r(\phi_{i} \to \phi_{f}) &\to r(\phi_{i} \to \phi_{f}), \\
\Delta^{w}_{Q}(g, \psi_{i}) &\to \Delta^{w}_{Q}(g, \psi_{i}) + gt, \\
\mathrm{Var}_{Q}(g, \psi_{i}) &\to \mathrm{Var}_{Q}(g, \psi_{i}).
\end{align}
}%
as $A \to A + t$.  From this,  one readily verifies that the uncertainty $\epsilon^{w}_{Q}(\eta;g,\psi_{i})$ is also invariant under translation. An analogous argument holds for $\epsilon^{w}_{P}(\eta;g,\psi_{i})$, and also for $\epsilon^{c}_{Q}(\eta;g,\psi_{i})$ in the conventional measurement model as well.

%%%%%%%%%%%%%%%%%%%%%%%%%%%%%%%%%%%%%%%%%%%%
%%%%%%%%%%%%%%%%%%%%%%
\section{Statistical Uncertainty in Weak Measurement}
%%%%%%%%%%%%%%%%%%%%%%
%%%%%%%%%%%%%%%%%%%%%%%%%%%%%%%%%%%%%%%%%%%%

In order to examine the statistical uncertainty in weak measurement, we take a closer look at the function
{\small
\begin{align}
\label{avpro}
{\Pi}^{N_0}_{X}(\kappa; g, \psi_{i}) 
:= \sum_{N=1}^{N_0} \mathrm{Bi}\left[N; N_0, r(\phi_{i}\to\phi_{f})\right]  \max \left[ \left( 1-\frac{\mathrm{Var}_{X}(\psi_{f})}{N\kappa^{2}} \right), 0 \right],
\end{align}
}%
where $\mathrm{Bi}\left[N; N_0,r\right] :=  \binom{N_0}{N} r^{N} \left( 1 - r \right)^{N_0-N}$ is the binomial distribution with $r$ given by the survival rate  \eqref{eq:sr}.
Since each summand is a continuous and monotonically increasing function in $\kappa$, the function ${\Pi}^{N_0}_{X}(\kappa; g, \psi_{i})$ inherits the same property. More explicitly, the function ${\Pi}^{N_0}_{X}(\kappa; g, \psi_{i})$ maps the interval $[0, \sqrt{\mathrm{Var}_{X}(\psi_{f})/N_{0}}]$ constantly to $0$ and is strictly monotonically increasing on $[\sqrt{\mathrm{Var}_{X}(\psi_{f})/N_{0}}, \infty)$ with $\lim_{\kappa \to \infty} {\Pi}^{N_0}_{X}(\kappa; g, \psi_{i}) = 1 - \mathrm{Bi}\left[0; N_0,r(\phi_{i}\to\phi_{f})\right]$. Hence, for $0 < \eta < 1 - \mathrm{Bi}\left[0; N_0,r(\phi_{i}\to\phi_{f})\right]$ the relation ${\Pi}^{N_0}_{X}(\kappa; g, \psi_{i}) = \eta$ can be solved in favor of $\kappa$
to obtain the inverse function $\kappa_Q^{N_0}(\eta; g, \psi_{i}) := [\Pi^{N_0}_{Q}]^{-1}(\eta; g, \psi_{i})$.
Now, since the postselection process is involved in weak measurement, it is always possible that \emph{zero} out of $N_{0}$ pairs of prepared states remains. This is the reason why the function $\kappa_Q^{N_0}(\eta; g, \psi_{i})$ can be defined only for $0<\eta <1 - \mathrm{Bi}\left[0; N_0,r(\phi_{i}\to\phi_{f})\right]$, in contrast to the conventional measurement case, where the choice $0 < \eta < 1$ is possible. Hence, we see that $\kappa_Q^{N_0}(\eta; g, \psi_{i})$ diverges to infinity as $\eta \to 1 - \mathrm{Bi}\left[0; N_0,r(\phi_{i}\to\phi_{f})\right]$, while its counterpart for the conventional measurement case remains finite. Based on this observation, we see that the postselection yields a larger statistical uncertainty for the weak measurement, which could quickly become uncontrollably large for higher $\eta$.

%%%%%%%%%%%%%%%%%%%%%%%%%%%%%%%%%%%%%%%%%%%%
%%%%%%%%%%%%%%%%%%%%%%
\section{Rewriting of the Transition Map}
%%%%%%%%%%%%%%%%%%%%%%
%%%%%%%%%%%%%%%%%%%%%%%%%%%%%%%%%%%%%%%%%%%%

For an observable with finite point spectrum, the transition maps $\{W_{g}\}_{g \in [0, \infty)}$ can be rewritten into a simple form. This fact proves to be useful in several aspects regarding weak measurement, in particular, for the proof of the existence of the limit of amplification, and analytic computation of the uncertainties, to name some, which we address later in this supplemental material.
Let
{\small
\begin{equation}
A = \sum_{n = 1}^{N} \lambda_{n} E_{\{\lambda_{n}\}}
\end{equation}
}%
be the spectral decomposition of $A$, where each $\lambda_{n}$ is an eigenvalue of $A$ and $E_{\{\lambda_{n}\}}$ is the projection on its accompanying eigenspace. Let $X: \mathcal{K} \supset \mathrm{dom}(X) \to \mathcal{K}$ be any observable on the meter coupled with $A$. The von Neumann type interaction is given by
{\small
\begin{align}
e^{-igA \otimes X} &= \sum_{k = 0}^{\infty} \frac{(-ig)^{k}}{k!} \left( A \otimes X \right)^{k} \nonumber \\
	&= \sum_{k = 0}^{\infty} \frac{(-ig)^{k}}{k!} \left( \left( \sum_{n = 1}^{N} \lambda_{n} E_{\{\lambda_{n}\}} \right)^{k} \otimes X^{k} \right) \nonumber \\
	&= \sum_{k = 0}^{\infty} \frac{(-ig)^{k}}{k!} \left( \left( \sum_{n = 1}^{N} \lambda_{n}^{k} E_{\{\lambda_{n}\}} \right) \otimes X^{k} \right) \nonumber \\
	&= \sum_{k = 0}^{\infty} \sum_{n = 1}^{N} \left( E_{\{\lambda_{n}\}} \otimes \frac{(-ig\lambda_{n})^{k}}{k!} X^{k} \right) \nonumber \\
	&= \sum_{n = 1}^{N} \left( E_{\{\lambda_{n}\}} \otimes \sum_{k = 0}^{\infty} \frac{(-ig\lambda_{n})^{k}}{k!} X^{k} \right) \nonumber \\
	&= \sum_{n = 1}^{N} \left( E_{\{\lambda_{n}\}} \otimes e^{-ig\lambda_{n}X} \right).
\end{align}
}%
Now, choose arbitrary preselected and postselected states of the system $| \phi_{i} \rangle, | \phi_{f} \rangle \in \mathcal{H}$, and arbitrary preselected state of the meter $| \psi_{i} \rangle \in \mathcal{K}$. One has
{\small
\begin{align}\label{eq:Transition_Operator_Finite}
| \psi_{f} \rangle &= \langle \phi_{f} | e^{-igA \otimes X} | \phi_{i} \rangle | \psi_{i} \rangle \nonumber \\
	&= \langle \phi_{f} | \sum_{n = 1}^{N} \left( E_{\{\lambda_{n}\}} \otimes e^{-ig\lambda_{n}X} \right) | \phi_{i} \rangle | \psi_{i} \rangle \nonumber \\
	&= \sum_{n = 1}^{N} \langle \phi_{f} | \left( E_{\{\lambda_{n}\}} \otimes e^{-ig\lambda_{n}X} \right) | \phi_{i} \rangle | \psi_{i} \rangle \nonumber \\
	&= \sum_{n = 1}^{N} \langle \phi_{f} | E_{\{\lambda_{n}\}} | \phi_{i} \rangle \cdot e^{-ig\lambda_{n}X} | \psi_{i} \rangle,
\end{align}
}%
and hence,
\begin{Lem}[Rewriting of the Transition Map]\label{Rewriting_the_Transition_Map}
In the setting above, for an observable  $A$ acting on  $\mathcal{H}$ with $1 \leq N < \infty$ point spectrum, the transition map is given in the form of
{\small
\begin{equation}\label{eq:Rewriting_the_Transition_Map}
W_{g} = \sum_{n = 1}^{N} \langle \phi_{f} | E_{\{\lambda_{n}\}} | \phi_{i} \rangle \cdot e^{-ig\lambda_{n}X},
\end{equation}
}%
where each $\lambda_{n}$ is an eigenvalue of $A$ and $E_{\{\lambda_{n}\}}$ is the projection on its accompanying eigenspace.
\end{Lem}

%%%%%%%%%%%%%%%%%%%%%%%%%%%%%%%%%%%%%%%%%%%%
%%%%%%%%%%%%%%%%%%%%%%
\section{Limit of Amplification of Shifts}
%%%%%%%%%%%%%%%%%%%%%%
%%%%%%%%%%%%%%%%%%%%%%%%%%%%%%%%%%%%%%%%%%%%

As a direct application of the result obtained in the previous section, we note that for an observable $A$ with finite eigenvalues, the shifts are bounded with respect to amplification of the weak value, whose fact is also mentioned in the last part of the  Letter. Observe from the above Lemma~\ref{Rewriting_the_Transition_Map} that, for fixed $| \psi_{i} \rangle \in \mathcal{K}$, the final state of the meter
{\small
\begin{equation}
| \psi_{f} \rangle = \sum_{n = 1}^{N} \langle \phi_{f} | E_{\{\lambda_{n}\}} | \phi_{i} \rangle \cdot e^{-ig\lambda_{n}X} | \psi_{i} \rangle
\end{equation}
}%
always lies in the subspace
{\small
\begin{equation}
| \psi_{f} \rangle \in \mathrm{lin} \left\{ e^{-ig\lambda_{1}X} | \psi_{i} \rangle, \dots, e^{-ig\lambda_{N}X} | \psi_{i} \rangle \right\}
\end{equation}
}%
of the meter $\mathcal{K}$, of dimension no greater than $N$, irrespective of the choice of $| \phi_{i} \rangle, | \phi_{f} \rangle \in \mathcal{H}$. Now, recall that any self-adjoint operator $X$ on a finite-dimensional normed space is necessarily bounded, and accordingly the collection $W(X) := \left\{ \langle\phi| X |\phi\rangle : \|\phi\|^{2} = 1 \right\}$ of all expectation values of $X$ is bounded.  We therefore have:
\begin{Prop}[Boundedness of the Shifts]\label{Boundedness_of_the_Shifts}
For an observable $A$ with finite eigenvalues, the shifts $\Delta^{w}_{X}(g, \psi_{i})$ ($X = Q, P$) are bounded with respect to any choice of preselection and postselection under fixed $g$ and $|\psi_{i}\rangle \in \mathcal{K}$.
In other words, the shifts cannot be amplified to an arbitrary extent only by the choice of preselection and postselection.
\end{Prop}

%%%%%%%%%%%%%%%%%%%%%%%%%%%%%%%%%%%%%%%%%%%%
%%%%%%%%%%%%%%%%%%%%%%
\section{Gaussian Model}
%%%%%%%%%%%%%%%%%%%%%%
%%%%%%%%%%%%%%%%%%%%%%%%%%%%%%%%%%%%%%%%%%%%

We now seek to construct an analytically computable model for the weak measurement, from which one can confirm the various statements made in the previous chapters and perform numerical estimations of the uncertainties as well.   In general, this entails a considerable difficulty in handling the transition map $W_{g} : |\psi_{i}\rangle \mapsto |\psi_{f}\rangle$ without resorting to approximations, but for the case in which the observable $A$ has finite eigenvalues, which we demonstrated above, the transition map can be rewritten into a tractable form.

Before embarking on the actual computation, we recall that the special case where the observable $A$ has the property $A^{2}=\mathrm{Id}$ has been studied earlier, for which we have 
\begin{equation}\label{eq:non_essential_expansion}
e^{-igA \otimes P} = \mathrm{Id} \otimes \cos{gP} -iA \otimes \sin{gP},
\end{equation}
so that $W_{g} = \langle\phi_{i} | \phi_{f}\rangle \left(\cos{gP} -i A_{\omega} \sin{gP}\right)$ for the transition map. The shifts for the Gaussian probe wave function was explicitly calculated based on this expansion \cite{Wu_2011, Koike_2011, Nakamura_2012}, revealing that the shifts cannot be amplified to an arbitrary extent by amplifying the weak value itself. 
Observing that an self-adjoint operator $A$ satisfies $A^{2}=\mathrm{Id}$ if and only if it has two eigenvalues $\{ -1, 1 \}$, we see that the condition $A^{2}=\mathrm{Id}$ is a special case to which we can apply Lemma~\ref{Rewriting_the_Transition_Map}. Indeed, based on \eqref{eq:Rewriting_the_Transition_Map} with $\{\lambda_{1}, \lambda_{2}\} = \{ -1, 1 \}$, we have
{\small
\begin{align}
W_{g} =& \langle \phi_{f} | E_{-1} | \phi_{i} \rangle e^{igP} + \langle \phi_{f} | E_{1} | \phi_{i} \rangle  e^{-igP} \nonumber \\
    =& \left(\langle \phi_{f} | E_{-1} | \phi_{i} \rangle + \langle \phi_{f} | E_{1} | \phi_{i} \rangle \right) \cos{gP} +i\left(\langle \phi_{f} | E_{-1} | \phi_{i} \rangle - \langle \phi_{f} | E_{1} | \phi_{i} \rangle \right) \sin{gP} \nonumber \\
	=& \langle \phi_{f} | \phi_{i} \rangle \left(\cos{gP} - iA_{\omega} \sin{gP}\right).
\end{align}
}%

In this respect, Lemma~\ref{Rewriting_the_Transition_Map} paves the way for the analytical investigation of the weak measurement model of a
more broader class of observables than was previously available. As for such demonstration, we slightly generalize the condition $A^{2}=\mathrm{Id}$ to the case in which $A$ has a discrete spectrum consisting of two distinct points $\{\lambda_{1}, \lambda_{2}\}$. For simplicity, we assume the Hilbert space of the meter to be $\mathcal{K} = L^{2}(\mathbb{R})$, and the collection of the meter's initial states $|\psi_{i}\rangle$ we can deploy are confined to normalized Gaussian wave functions
{\small
\begin{equation}
 \psi_{i}(x) = \left( \frac{1}{\pi d^{2}} \right)^{\frac{1}{4}} e^{-\frac{x^{2}}{2d^{2}}},
\end{equation}
}%
centered at $x = 0$ with width $d > 0$.

Under this condition, we seek to obtain formulae for the quantities required for evaluating the uncertainties, namely, the survival rate $r(\phi_{i} \to \phi_{f})$, the shifts $\Delta^{w}_{X}(g, \psi_{i})$ and the variances $\mathrm{Var}_{X}(\psi_{f})$ for each $X = Q,P$.
Let
$A = \lambda_{1}E_{1} + \lambda_{2}E_{2}$
be the spectral decomposition of $A$, where $\lambda_{n}$, $n=1,2$, are its two discrete eigenvalues, and $E_{n}$, $n=1, 2$, are the projection operators onto their respective eigenspaces. By defining $c_{n} := \langle\phi_{f}| E_{n} |\phi_{i}\rangle$ for arbitrary normalized non-orthogonal pair of $|\phi_{i}\rangle, |\phi_{f}\rangle$, we see that the final state of the meter can be rewritten as
\begin{align}
\psi_{f}(x) &= c_{1} \cdot e^{-ig\lambda_{1}\hat{p}} \psi_{i}(x) + c_{2} \cdot e^{-ig\lambda_{2}\hat{p}} \psi_{i}(x) \nonumber \\
    &= c_{1} \psi_{i}(x - g\lambda_{1}) + c_{2} \psi_{i}(x - g\lambda_{2}),
\end{align}
where, for this Gaussian case, we have $\psi_{i}(x - g\lambda_{n}) = \left( 1/\pi d^{2} \right)^{\frac{1}{4}} e^{-(x - g\lambda_{n})^{2}/2d^{2}}$.  For convenience, we also introduce
the shorthands $\Lambda_{m} := (\lambda_{1} + \lambda_{2})/{2}$, $\Lambda_{r} := (\lambda_{2} - \lambda_{1})/{2}$ and
{\small
\begin{gather}
A_{r} := A_{w} - \Lambda_{m},\\
a := \frac{1}{2} \left(\frac{\vert A_{r} \vert ^{2}}{\Lambda_{r}^{2}} - 1\right),
\end{gather}
}%
for the weak value $A_{w}$ in \eqref{eq:wvalue}.
Since we have
$\langle\phi_{f}| A |\phi_{i}\rangle = \lambda_{1}c_{1} + \lambda_{2}c_{2}$ and $\langle\phi_{f} | \phi_{i}\rangle = c_{1} + c_{2}$, the above quantities can be cast into
{\small
\begin{gather}
	\mathrm{Re}A_{r} = \Lambda_{r} \cdot \frac{- |c_{1}|^{2} + |c_{2}|^{2}}{|c_{1}|^{2} + |c_{2}|^{2} + 2\mathrm{Re}\left[c_{1}^{*}c_{2}\right]}, \\
	\mathrm{Im}A_{r} = \Lambda_{r} \cdot \frac{2 \mathrm{Im}\left[c_{1}^{*}c_{2}\right] }{|c_{1}|^{2} + |c_{2}|^{2} + 2\mathrm{Re}\left[c_{1}^{*}c_{2}\right]}, \\
	a = - \frac{2\mathrm{Re} \left[c_{1}^{*}c_{2}\right]}{\lvert c_{1} \rvert^{2} + \lvert c_{2} \rvert^{2} + 2\mathrm{Re} \left[c_{1}^{*}c_{2}\right]},
\end{gather}
}%
in terms of $\{c_{n}\}$. We also note the following formulae for later convenience:
{\small
\begin{gather}
\int_{\mathbb{R}} e^{-(x - g\lambda_{m})^{2}/2d^{2}} \cdot e^{-(x - g\lambda_{n})^{2}/2d^{2}} dx = \left( \pi d^{2} \right)^{\frac{1}{2}} \exp \left[-\frac{g^{2}}{d^{2}} \left( \frac{\lambda_{m} - \lambda_{n}}{2} \right)^{2} \right], \\
    \int_{\mathbb{R}} e^{-(x - g\lambda_{m})^{2}/2d^{2}} \cdot x e^{-(x - g\lambda_{n})^{2}/2d^{2}} dx = \left( \pi d^{2} \right)^{\frac{1}{2}} \left(g \cdot \frac{\lambda_{m} + \lambda_{n}}{2} \right) \exp \left[-\frac{g^{2}}{d^{2}} \left( \frac{\lambda_{m} - \lambda_{n}}{2} \right)^{2} \right], \\
    \int_{\mathbb{R}} e^{-(x - g\lambda_{m})^{2}/2d^{2}} \cdot x^{2} e^{-(x - g\lambda_{n})^{2}/2d^{2}} dx = \left( \pi d^{2} \right)^{\frac{1}{2}} \left( \frac{d^{2}}{2} + \left( g \cdot \frac{\lambda_{m} + \lambda_{n}}{2} \right)^{2} \right) \exp \left[-\frac{g^{2}}{d^{2}} \left( \frac{\lambda_{m} - \lambda_{n}}{2} \right)^{2} \right], \\
    \int_{\mathbb{R}} e^{-(x - g\lambda_{m})^{2}/2d^{2}} \cdot \left(-i \frac{d}{dx} \right) e^{-(x - g\lambda_{n})^{2}/2d^{2}} dx = \left( \pi d^{2} \right)^{\frac{1}{2}} \left( i\frac{g}{d^{2}} \cdot \frac{\lambda_{m} - \lambda_{n}}{2} \right) \exp \left[-\frac{g^{2}}{d^{2}} \left( \frac{\lambda_{m} - \lambda_{n}}{2} \right)^{2} \right], \\
    \int_{\mathbb{R}} e^{-(x - g\lambda_{m})^{2}/2d^{2}} \cdot \left(-i \frac{d}{dx} \right)^{2} e^{-(x - g\lambda_{n})^{2}/2d^{2}} dx = \left( \pi d^{2} \right)^{\frac{1}{2}} \left( \frac{1}{2d^{2}} - \left( \frac{g}{d^{2}} \cdot \frac{\lambda_{m} - \lambda_{n}}{2} \right)^{2} \right) \exp \left[-\frac{g^{2}}{d^{2}} \left( \frac{\lambda_{m} - \lambda_{n}}{2} \right)^{2} \right].
\end{gather}
}%

Armed with these formulae, we can explicitly evaluate the survival rate, the shifts of the meter positions and their variaces, which we list one by one below.

%%%%%%%%%%%%%%%%%%%%%%
\subsection{Survival Rate}
%%%%%%%%%%%%%%%%%%%%%%

\begin{Prop}[Survival Rate]
The survival rate $r(\phi_{i} \to \phi_{f})$ is given by
{\small
\begin{equation}\label{eq:sr_gauss}
r(\phi_{i} \to \phi_{f}) = \left|\langle\phi_{f} | \phi_{i}\rangle\right|^{2}\left[1 + a \left( 1 - e^{-g^{2}\Lambda_{r}^{2}/d^{2}} \right)\right].
\end{equation}
}%
\end{Prop}
\begin{Prf}
Since all $|\phi_{i}\rangle, |\phi_{f}\rangle \in \mathcal{H}$ and $|\phi_{i}\rangle \in \mathcal{K}$ are assumed to be normalized here, by \eqref{eq:sr} one has
{\small
\begin{align}
r(\phi_{i} \to \phi_{f}) &= \|\psi_{f}\|_{L^{2}(\mathbb{R})}^{2} \nonumber \\
    &= \left( \lvert c_{1} \rvert^{2} + \lvert c_{2} \rvert^{2} \right) \|\psi_{i}\|_{L^{2}(\mathbb{R})}^{2} + 2\mathrm{Re} \left[c_{1}^{*}c_{2} \langle e^{-ig\lambda_{1}\hat{p}}\psi_{i} \mid e^{-ig\lambda_{2}\hat{p}} \psi_{i}\rangle \right].
\end{align}
}%
For $\psi_{i}(x) = ( 1/\pi d^{2} )^{\frac{1}{4}} e^{-\frac{x^{2}}{2d^{2}}}$, one has
{\small
\begin{align}
\langle e^{-ig\lambda_{1}\hat{p}}\psi_{i} \mid e^{-ig\lambda_{2}\hat{p}} \psi_{i}\rangle &:= \int_{\mathbb{R}} \psi_{i}(x - g\lambda_{1})\psi_{i}(x - g\lambda_{2}) dx \nonumber \\
	&= \left( \frac{1}{\pi d^{2}} \right)^{\frac{1}{2}} \int_{\mathbb{R}} e^{-(x - g\lambda_{1})^{2}/2d^{2}} e^{-(x - g\lambda_{2})^{2}/2d^{2}} dx \nonumber \\
	&= e^{-g^{2}\Lambda_{r}^{2}/d^{2}},
\end{align}
}%
hence
{\small
\begin{align}\label{denominator}
\|\psi_{f}\|_{L^{2}(\mathbb{R})}^{2} &= \lvert c_{1} \rvert^{2} + \lvert c_{2} \rvert^{2} + 2\mathrm{Re} \left[c_{1}^{*}c_{2}\right] e^{-g^{2}\Lambda_{r}^{2}/d^{2}} \nonumber \\
	&= \left( \lvert c_{1} \rvert^{2} + \lvert c_{2} \rvert^{2} + 2\mathrm{Re} \left[c_{1}^{*}c_{2}\right] \right) - 2\mathrm{Re} \left[c_{1}^{*}c_{2}\right] \left( 1 - e^{-g^{2}\Lambda_{r}^{2}/d^{2}} \right) \nonumber \\
	&= \left|\langle\phi_{f} | \phi_{i}\rangle\right|^{2}\left[1 + a \left( 1 - e^{-g^{2}\Lambda_{r}^{2}/d^{2}} \right)\right].
\end{align}
}%
\qed
\end{Prf}

%%%%%%%%%%%%%%%%%%%%%%
\subsection{Shifts}
%%%%%%%%%%%%%%%%%%%%%%

\begin{Prop}[Shift of the Position]
The shift of the position $\hat{x}$ is given by
{\small
\begin{equation}\label{eq:shift_x_gauss}
\Delta_{\hat{x}}^{w}(g,d) = g \cdot \frac{\mathrm{Re} A_{r}}{1 + a\left( 1 - e^{-g^{2}\Lambda_{r}^{2}/d^{2}} \right)} + g \cdot \Lambda_{m}.
\end{equation}
}%
\end{Prop}

\begin{Prf}
Observe that
{\small
\begin{align}
\langle\psi_{f}| \hat{x} |\psi_{f}\rangle =& \lvert c_{1} \rvert^{2} \langle e^{-ig\lambda_{1}\hat{p}}\psi_{i} | \hat{x} | e^{-ig\lambda_{1}\hat{p}} \psi_{i}\rangle + \lvert c_{2} \rvert^{2} \langle e^{-ig\lambda_{2}\hat{p}}\psi_{i} | \hat{x} | e^{-ig\lambda_{2}\hat{p}} \psi_{i}\rangle \nonumber \\
	&+ 2\mathrm{Re} \left[c_{1}^{*}c_{2} \langle e^{-ig\lambda_{1}\hat{p}}\psi_{i} | \hat{x} | e^{-ig\lambda_{2}\hat{p}} \psi_{i}\rangle \right].
\end{align}
}%
For $\psi_{i}(x) = ( 1/\pi d^{2} )^{\frac{1}{4}} e^{-\frac{x^{2}}{2d^{2}}}$, one has
{\small
\begin{align}
\langle e^{-ig\lambda_{m}\hat{p}}\psi_{i} | \hat{x} | e^{-ig\lambda_{n}\hat{p}} \psi_{i}\rangle &:= \left( \frac{1}{\pi d^{2}} \right)^{\frac{1}{2}} \int_{\mathbb{R}} x e^{-(x - g\lambda_{m})^{2}/2d^{2}} e^{-(x - g\lambda_{n})^{2}/2d^{2}} dx \nonumber \\
	&= g \cdot \frac{\lambda_{m} + \lambda_{n}}{2} \cdot e^{-\frac{g^{2}}{d^{2}}\left(\frac{\lambda_{m} - \lambda_{n}}{2}\right)^{2}},
\end{align}
}%
hence
{\small
\begin{align}
\langle\psi_{f}| \hat{x} |\psi_{f}\rangle &= g \left( \lvert c_{1} \rvert^{2}\lambda_{1} + \lvert c_{2} \rvert^{2}\lambda_{2} \right) + 2 g \Lambda_{m} \mathrm{Re} \left[c_{1}^{*}c_{2}\right] e^{-g^{2}\Lambda_{r}^{2}/d^{2}}.
\end{align}
}%
Now, since $E_{\hat{x}}(\psi_{i}) = 0$, one has
{\small
\begin{align}
\Delta_{\hat{x}}^{w}(g,d) &:= E_{\hat{x}}(\psi_{f}) - E_{\hat{x}}(\psi_{i})\nonumber \\
    &= g \cdot \frac{\left( \lvert c_{1} \rvert^{2}\lambda_{1} + \lvert c_{2} \rvert^{2}\lambda_{2} \right) + 2\Lambda_{m}\mathrm{Re} \left[c_{1}^{*}c_{2}\right] e^{-g^{2}\Lambda_{r}^{2}/d^{2}}}{\lvert c_{1} \rvert^{2} + \lvert c_{2} \rvert^{2} + 2\mathrm{Re} \left[c_{1}^{*}c_{2}\right] e^{-g^{2}\Lambda_{r}^{2}/d^{2}}} \nonumber \\
    &= g \cdot \frac{ \Lambda_{r} \cdot \left( - \lvert c_{1} \rvert^{2}\lambda_{1} + \lvert c_{2} \rvert^{2}\lambda_{2} \right) }{\lvert c_{1} \rvert^{2} + \lvert c_{2} \rvert^{2} + 2\mathrm{Re} \left[c_{1}^{*}c_{2}\right] e^{-g^{2}\Lambda_{r}^{2}/d^{2}}} + g \cdot \Lambda_{m} \nonumber \\
    &= g \cdot \frac{ \Lambda_{r} \cdot \left( - \lvert c_{1} \rvert^{2}\lambda_{1} + \lvert c_{2} \rvert^{2}\lambda_{2} \right) }{\ \left( \lvert c_{1} \rvert^{2} + \lvert c_{2} \rvert^{2} + 2\mathrm{Re} \left[c_{1}^{*}c_{2}\right] \right) - 2\mathrm{Re} \left[c_{1}^{*}c_{2}\right] \left( 1 - e^{-g^{2}\Lambda_{r}^{2}/d^{2}} \right) } + g \cdot \Lambda_{m} \nonumber \\
	&= g \cdot \frac{\mathrm{Re} A_{r}}{1 + a\left( 1 - e^{-g^{2}\Lambda_{r}^{2}/d^{2}} \right)} + g \cdot \Lambda_{m}.
\end{align}
}%
\qed
\end{Prf}

\begin{Prop}[Shift of the Momentum]
The shift of the momentum $\hat{p}$ is given by
{\small
\begin{equation}\label{eq:shift_p_gauss}
\Delta_{\hat{p}}^{w}(g,d) = \frac{g}{d^{2}} \cdot \frac{ \mathrm{Im} A_{r} e^{-2g^{2}W^{2}\Lambda_{r}^{2}}}{1 + a\left( 1 - e^{-g^{2}\Lambda_{r}^{2}/d^{2}} \right)}.
\end{equation}
}%
\end{Prop}

\begin{Prf}
Observe that
{\small
\begin{align}
\langle\psi_{f}| \hat{p} |\psi_{f}\rangle =& \lvert c_{1} \rvert^{2} \langle e^{-ig\lambda_{1}\hat{p}}\psi_{i} | \hat{p} | e^{-ig\lambda_{1}\hat{p}} \psi_{i}\rangle + \lvert c_{2} \rvert^{2} \langle e^{-ig\lambda_{2}\hat{p}}\psi_{i} | \hat{p} | e^{-ig\lambda_{2}\hat{p}} \psi_{i}\rangle \nonumber \\
	&+ 2\mathrm{Re} \left[c_{1}^{*}c_{2} \langle e^{-ig\lambda_{1}\hat{p}}\psi_{i} | \hat{p} | e^{-ig\lambda_{2}\hat{p}} \psi_{i}\rangle \right].
\end{align}
}%
For $\psi_{i}(x) = ( 1/\pi d^{2} )^{\frac{1}{4}} e^{-\frac{x^{2}}{2d^{2}}}$, one has
{\small
\begin{align}
\langle e^{-ig\lambda_{i}\hat{p}}\psi_{i} | \hat{p} | e^{-ig\lambda_{j}\hat{p}} \psi_{i}\rangle &:= \left( \frac{1}{\pi d^{2}} \right)^{\frac{1}{2}} \int_{\mathbb{R}} e^{-(x - g\lambda_{m})^{2}/2d^{2}} \cdot \left(-i \frac{d}{dx} \right) e^{-(x - g\lambda_{n})^{2}/2d^{2}} dx \nonumber \\
	&= i\frac{g}{d^{2}} \cdot \frac{\lambda_{m} - \lambda_{n}}{2} \cdot  \exp \left[-\frac{g^{2}}{d^{2}} \left( \frac{\lambda_{m} - \lambda_{n}}{2} \right)^{2} \right],
\end{align}
}%
hence
{\small
\begin{align}
\langle\psi_{f}| \hat{p} |\psi_{f}\rangle &= 2 \frac{g}{d^{2}} \Lambda_{r} \mathrm{Im} \left[c_{1}^{*}c_{2}\right] e^{-g^{2}\Lambda_{r}^{2}/d^{2}}.
\end{align}
}%
Now, since $E_{\hat{p}}(\psi_{i}) = 0$, one has
{\small
\begin{align}
\Delta_{\hat{p}}^{w}(g,d) &:= E_{\hat{p}}(\psi_{f}) - E_{\hat{p}}(\psi_{i})\nonumber \\
    &= \frac{g}{d^{2}} \cdot \frac{ 2\Lambda_{r} \mathrm{Im} \left[c_{1}^{*}c_{2}\right]  e^{-g^{2}\Lambda_{r}^{2}/d^{2}} }{\lvert c_{1} \rvert^{2} + \lvert c_{2} \rvert^{2} + 2\mathrm{Re} \left[c_{1}^{*}c_{2}\right] e^{-g^{2}\Lambda_{r}^{2}/d^{2}}} \nonumber \\
	&= \frac{g}{d^{2}} \cdot \frac{ 2 \Lambda_{r} \mathrm{Im} \left[c_{1}^{*}c_{2}\right]  e^{-g^{2}\Lambda_{r}^{2}/d^{2}} }{\left( \lvert c_{1} \rvert^{2} + \lvert c_{2} \rvert^{2} + 2\mathrm{Re} \left[c_{1}^{*}c_{2}\right] \right) - 2\mathrm{Re} \left[c_{1}^{*}c_{2}\right] \left( 1 - e^{-g^{2}\Lambda_{r}^{2}/d^{2}} \right)} \nonumber \\
	&= \frac{g}{d^{2}} \cdot \frac{ \mathrm{Im} A_{r} e^{-2g^{2}W^{2}\Lambda_{r}^{2}}}{1 + a\left( 1 - e^{-g^{2}\Lambda_{r}^{2}/d^{2}} \right)}.
\end{align}
}%
\qed
\end{Prf}

%%%%%%%%%%%%%%%%%%%%%%
\subsection{Variances}
%%%%%%%%%%%%%%%%%%%%%%

\begin{Lem}[Position Variance]
The variance of the position $\hat{x}$ on the state $\psi_{f}$ is given by
{\small
\begin{equation}\label{eq:var_x_gauss}
\mathrm{Var}_{\hat{x}}(\psi_{f}) = \frac{d^{2}}{2} + g^{2} \cdot \frac{  \Lambda_{r}^{2} \left( 1 + a \right) }{1 + a\left( 1 - e^{-g^{2}\Lambda_{r}^{2}/d^{2}} \right)}  - g^{2} \cdot\left(  \frac{\mathrm{Re} A_{r}}{1 + a\left( 1 - e^{-g^{2}\Lambda_{r}^{2}/d^{2}} \right)} \right)^{2}.
\end{equation}
}%
\end{Lem}

\begin{Prf}
Observe that
{\small
\begin{align}
\langle\psi_{f}| \hat{x}^{2} |\psi_{f}\rangle =& \lvert c_{1} \rvert^{2} \langle e^{-ig\lambda_{1}\hat{p}}\psi_{i} | \hat{x}^{2} | e^{-ig\lambda_{1}\hat{p}} \psi_{i}\rangle + \lvert c_{2} \rvert^{2} \langle e^{-ig\lambda_{2}\hat{p}}\psi_{i} | \hat{x}^{2} | e^{-ig\lambda_{2}\hat{p}} \psi_{i}\rangle \nonumber \\
	&+ 2\mathrm{Re} \left[c_{1}^{*}c_{2} \langle e^{-ig\lambda_{1}\hat{p}}\psi_{i} | \hat{x}^{2} | e^{-ig\lambda_{2}\hat{p}} \psi_{i}\rangle \right].
\end{align}
}%
For $\psi_{i}(x) = ( 1/\pi d^{2} )^{\frac{1}{4}} e^{-\frac{x^{2}}{2d^{2}}}$, one has
{\small
\begin{align}
\langle e^{-ig\lambda_{1}\hat{p}}\psi_{i} | \hat{x}^{2} | e^{-ig\lambda_{2}\hat{p}} \psi_{i}\rangle &:= \left( \frac{1}{\pi d^{2}} \right)^{\frac{1}{2}}  \int_{\mathbb{R}} e^{-(x - g\lambda_{m})^{2}/2d^{2}} \cdot x^{2} e^{-(x - g\lambda_{n})^{2}/2d^{2}} dx \nonumber \\
    &= \left( \frac{d^{2}}{2} + \left( g \cdot \frac{\lambda_{m} + \lambda_{n}}{2} \right)^{2} \right) \exp \left[-\frac{g^{2}}{d^{2}} \left( \frac{\lambda_{m} - \lambda_{n}}{2} \right)^{2} \right].
\end{align}
}%
hence
{\small
\begin{equation}
\langle\psi_{f}| \hat{x}^{2} |\psi_{f}\rangle = \frac{d^{2}}{2} \left( \lvert c_{1} \rvert^{2} + \lvert c_{2} \rvert^{2} + 2 \mathrm{Re} \left[c_{1}^{*}c_{2}\right] e^{-g^{2}\Lambda_{r}^{2}/d^{2}} \right) + g^{2} \left( \lambda_{1}^{2}\lvert c_{1} \rvert^{2} + \lambda_{2}^{2}\lvert c_{2} \rvert^{2} +  2 \Lambda_{m}^{2} \mathrm{Re} \left[c_{1}^{*}c_{2}\right] e^{-g^{2}\Lambda_{r}^{2}/d^{2}} \right) ,
\end{equation}
}%
which leads to
{\small
\begin{align}
E_{\hat{x}^{2}}(\psi_{f}) &= \frac{\frac{d^{2}}{2} \left( \lvert c_{1} \rvert^{2} + \lvert c_{2} \rvert^{2} + 2 \mathrm{Re} \left[c_{1}^{*}c_{2}\right] e^{-g^{2}\Lambda_{r}^{2}/d^{2}} \right) + g^{2} \left( \lambda_{1}^{2}\lvert c_{1} \rvert^{2} + \lambda_{2}^{2}\lvert c_{2} \rvert^{2} +  2 \Lambda_{m}^{2} \mathrm{Re} \left[c_{1}^{*}c_{2}\right] e^{-g^{2}\Lambda_{r}^{2}/d^{2}} \right)}{\lvert c_{1} \rvert^{2} + \lvert c_{2} \rvert^{2} + 2\mathrm{Re} \left[c_{1}^{*}c_{2}\right] e^{-g^{2}\Lambda_{r}^{2}/d^{2}}} \nonumber \\
    &= \frac{d^{2}}{2} + g^{2} \cdot \frac{\left( \lambda_{1}^{2}\lvert c_{1} \rvert^{2} + \lambda_{2}^{2}\lvert c_{2} \rvert^{2} \right) - \Lambda_{m}^{2} \left( \lvert c_{1} \rvert^{2} + \lvert c_{2} \rvert^{2} \right) }{\lvert c_{1} \rvert^{2} + \lvert c_{2} \rvert^{2} + 2\mathrm{Re} \left[c_{1}^{*}c_{2}\right] e^{-g^{2}\Lambda_{r}^{2}/d^{2}}} + g^{2} \Lambda_{m}^{2} \nonumber \\
    &= \frac{d^{2}}{2} + g^{2} \cdot \frac{\left( \lambda_{1}^{2}\lvert c_{1} \rvert^{2} + \lambda_{2}^{2}\lvert c_{2} \rvert^{2} + 2\lambda_{1}\lambda_{2}\mathrm{Re} \left[c_{1}^{*}c_{2}\right] \right) - 2\lambda_{1}\lambda_{2}\mathrm{Re} \left[c_{1}^{*}c_{2}\right] - \Lambda_{m}^{2} \left( \lvert c_{1} \rvert^{2} + \lvert c_{2} \rvert^{2} \right) }{\left( \lvert c_{1} \rvert^{2} + \lvert c_{2} \rvert^{2} + 2\mathrm{Re} \left[c_{1}^{*}c_{2}\right] \right) - 2\mathrm{Re} \left[c_{1}^{*}c_{2}\right] \left( 1 - e^{-g^{2}\Lambda_{r}^{2}/d^{2}} \right)} + g^{2} \Lambda_{m}^{2} \nonumber \\
    &= \frac{d^{2}}{2} + g^{2} \cdot \frac{\lvert A_{r} + \Lambda_{m} \rvert^{2} + \lambda_{1}\lambda_{2}a - \Lambda_{m}^{2} \left( 1+a \right) }{1 + a\left( 1 - e^{-g^{2}\Lambda_{r}^{2}/d^{2}} \right)} + g^{2} \Lambda_{m}^{2} \nonumber \\
    &= \frac{d^{2}}{2} + g^{2} \cdot \frac{\lvert A_{r} \rvert^{2} + 2\Lambda_{m} \mathrm{Re}A_{r} - \Lambda_{r}^{2} a }{1 + a\left( 1 - e^{-g^{2}\Lambda_{r}^{2}/d^{2}} \right)} + g^{2} \Lambda_{m}^{2}.
\end{align}
}%
From the above results, one obtains
{\small
\begin{align}
\mathrm{Var}_{\hat{x}}(\psi_{f}) := & E_{\hat{x}^{2}}(\psi_{f}) - E_{\hat{x}}(\psi_{f})^{2} \nonumber \\
	=& \frac{d^{2}}{2} + g^{2} \cdot \frac{\lvert A_{r} \rvert^{2} + 2\Lambda_{m} \mathrm{Re} A_{r} - \Lambda_{r}^{2} a }{1 + a\left( 1 - e^{-g^{2}\Lambda_{r}^{2}/d^{2}} \right)} + g^{2} \Lambda_{m}^{2} - \left( g \cdot \frac{\mathrm{Re} A_{r}}{1 + a\left( 1 - e^{-g^{2}\Lambda_{r}^{2}/d^{2}} \right)} + g \cdot \Lambda_{m} \right)^{2} \nonumber \\
	=& \frac{d^{2}}{2} + g^{2} \cdot \frac{\lvert A_{r} \rvert^{2} - \Lambda_{r}^{2} a }{1 + a\left( 1 - e^{-g^{2}\Lambda_{r}^{2}/d^{2}} \right)}  - g^{2} \cdot\left(  \frac{\mathrm{Re} A_{r}}{1 + a\left( 1 - e^{-g^{2}\Lambda_{r}^{2}/d^{2}} \right)} \right)^{2} \nonumber \\
	=& \frac{d^{2}}{2} + g^{2} \cdot \frac{  \Lambda_{r}^{2} \left( 1 + a \right) }{1 + a\left( 1 - e^{-g^{2}\Lambda_{r}^{2}/d^{2}} \right)}  - g^{2} \cdot\left(  \frac{\mathrm{Re} A_{r}}{1 + a\left( 1 - e^{-g^{2}\Lambda_{r}^{2}/d^{2}} \right)} \right)^{2}.
\end{align}
}%
\qed
\end{Prf}

\begin{Prop}[Momentum Variance]
The variance of the momentum $\hat{p}$ on the state $\psi_{f}$ is given by
{\small
\begin{equation}\label{eq:var_p_gauss}
\mathrm{Var}_{\hat{p}}(\psi_{f}) = \frac{1}{2d^{2}} + \left(\frac{g}{d^{2}}\right)^{2} \left[ \frac{ a \Lambda_{r}^{2} e^{-g^{2}\Lambda_{r}^{2}/d^{2}} }{1 + a\left( 1 - e^{-g^{2}\Lambda_{r}^{2}/d^{2}} \right)} - \left( \frac{ \mathrm{Im} A_{r} e^{-g^{2}\Lambda_{r}^{2}/d^{2}}}{1 + a\left( 1 - e^{-g^{2}\Lambda_{r}^{2}/d^{2}} \right)} \right)^{2} \right].
\end{equation}
}%
\end{Prop}

\begin{Prf}
Observe that
{\small
\begin{align}
\langle\psi_{f}| \hat{p}^{2} |\psi_{f}\rangle &= \lvert c_{1} \rvert^{2} \langle e^{-ig\lambda_{1}\hat{p}}\psi_{i} | \hat{p}^{2} | e^{-ig\lambda_{1}\hat{p}} \psi_{i}\rangle + \lvert c_{2} \rvert^{2} \langle e^{-ig\lambda_{2}\hat{p}}\psi_{i} | \hat{p}^{2} | e^{-ig\lambda_{2}\hat{p}} \psi_{i}\rangle \nonumber \\
	&+ 2\mathrm{Re} \left[c_{1}^{*}c_{2} \langle e^{-ig\lambda_{1}\hat{p}}\psi_{i} | \hat{p}^{2} | e^{-ig\lambda_{2}\hat{p}} \psi_{i}\rangle \right].
\end{align}
}%
For $\psi_{i}(x) = ( 1/\pi d^{2} )^{\frac{1}{4}} e^{-\frac{x^{2}}{2d^{2}}}$, one has
{\small
\begin{align}
\langle e^{-ig\lambda_{1}\hat{p}}\psi_{i} | \hat{p}^{2} | e^{-ig\lambda_{2}\hat{p}} \psi_{i}\rangle &:= \left( \frac{1}{\pi d^{2}} \right)^{\frac{1}{2}} \int_{\mathbb{R}} e^{-(x - g\lambda_{m})^{2}/2d^{2}} \cdot \left(-i \frac{d}{dx} \right)^{2} e^{-(x - g\lambda_{n})^{2}/2d^{2}} dx \nonumber \\
    &= \left( \frac{1}{2d^{2}} - \left( \frac{g}{d^{2}} \cdot \frac{\lambda_{m} - \lambda_{n}}{2} \right)^{2} \right) \exp \left[-\frac{g^{2}}{d^{2}} \left( \frac{\lambda_{m} - \lambda_{n}}{2} \right)^{2} \right],
\end{align}
}%
hence
{\small
\begin{equation}
\langle\psi_{f}| \hat{p}^{2} |\psi_{f}\rangle = \frac{1}{2d^{2}} \left(  \lvert c_{1} \rvert^{2} +  \lvert c_{2} \rvert^{2} + 2\mathrm{Re} \left[c_{1}^{*}c_{2}\right]e^{-g^{2}\Lambda_{r}^{2}/d^{2}} \right) - \left(\frac{g}{d^{2}}\right)^{2}\Lambda_{r}^{2}2\mathrm{Re} \left[c_{1}^{*}c_{2}\right]e^{-g^{2}\Lambda_{r}^{2}/d^{2}},
\end{equation}
}%
which leads to
{\small
\begin{align}
E_{\hat{p}^{2}}(\psi_{f}) &= \frac{\frac{1}{2d^{2}} \left(  \lvert c_{1} \rvert^{2} +  \lvert c_{2} \rvert^{2} + 2\mathrm{Re} \left[c_{1}^{*}c_{2}\right]e^{-g^{2}\Lambda_{r}^{2}/d^{2}} \right) - \left(\frac{g}{d^{2}}\right)^{2}\Lambda_{r}^{2}2\mathrm{Re} \left[c_{1}^{*}c_{2}\right]e^{-g^{2}\Lambda_{r}^{2}/d^{2}} }{\lvert c_{1} \rvert^{2} + \lvert c_{2} \rvert^{2} + 2\mathrm{Re} \left[c_{1}^{*}c_{2}\right] e^{-g^{2}\Lambda_{r}^{2}/d^{2}}} \nonumber \\
	&= \frac{1}{2d^{2}} - \left(\frac{g}{d^{2}}\right)^{2} \cdot \frac{ \Lambda_{r}^{2} 2\mathrm{Re} \left[c_{1}^{*}c_{2}\right] e^{-g^{2}\Lambda_{r}^{2}/d^{2}} }{\left( \lvert c_{1} \rvert^{2} + \lvert c_{2} \rvert^{2} + 2\mathrm{Re} \left[c_{1}^{*}c_{2}\right] \right) - 2\mathrm{Re} \left[c_{1}^{*}c_{2}\right] \left( 1 - e^{-g^{2}\Lambda_{r}^{2}/d^{2}} \right)} \nonumber \\
	&= \frac{1}{2d^{2}} + \left(\frac{g}{d^{2}}\right)^{2} \cdot \frac{ a \Lambda_{r}^{2} e^{-g^{2}\Lambda_{r}^{2}/d^{2}} }{1 + a\left( 1 - e^{-g^{2}\Lambda_{r}^{2}/d^{2}} \right)}.
\end{align}
}%
One thus finds
{\small
\begin{align}
\mathrm{Var}_{\hat{p}}(\psi_{f}) &:= E_{\hat{p}^{2}}(\psi_{f}) - E_{\hat{p}}(\psi_{f})^{2} \nonumber \\
    &= \frac{1}{2d^{2}} + \left(\frac{g}{d^{2}}\right)^{2} \left[ \frac{ a \Lambda_{r}^{2} e^{-g^{2}\Lambda_{r}^{2}/d^{2}} }{1 + a\left( 1 - e^{-g^{2}\Lambda_{r}^{2}/d^{2}} \right)} - \left( \frac{ \mathrm{Im} A_{r} e^{-g^{2}\Lambda_{r}^{2}/d^{2}}}{1 + a\left( 1 - e^{-g^{2}\Lambda_{r}^{2}/d^{2}} \right)} \right)^{2} \right].
\end{align}
}%
\qed
\end{Prf}

%%%%%%%%%%%%%%%%%%%%%%
\subsection{Summary: Total Uncertainties}
%%%%%%%%%%%%%%%%%%%%%%

Summing up, we have so far obtained quantities necessary for the analytical computation of the uncertainties, namely, the survival rate \eqref{eq:sr_gauss}, the shifts \eqref{eq:shift_x_gauss}, \eqref{eq:shift_p_gauss} and the variances \eqref{eq:var_x_gauss}, \eqref{eq:var_p_gauss} under the simple model in which the observable $A$ of the system has a discrete spectrum consisting of two distinct points $\{\lambda_{1}, \lambda_{2}\}$, and the initial states of the meter are confined to normalized Gaussian wave functions in $L^{2}(\mathbb{R})$ centered at $x = 0$.
Based on the above results and formula (13) (with its counterpart for $P$) in the Letter, we finally arrive at the desired total uncertainties:
{\small
\begin{gather}
\epsilon^{w}_{\hat{x}}(\eta; g, d) = \frac{\delta_{\hat{x}}}{g} + \frac{\kappa^{N_{0}}_{\hat{x}}(\eta;g,d)}{g} + \left\vert \mathrm{Re} A_{r} \right\vert \cdot \left\vert \frac{a\left( 1 - e^{-g^{2}\Lambda_{r}^{2}/d^{2}} \right)}{1 + a\left( 1 - e^{-g^{2}\Lambda_{r}^{2}/d^{2}} \right)} \right\vert, \label{eq:uncert_x_gauss} \\
\epsilon^{w}_{\hat{p}}(\eta; g, d) = \frac{\delta_{\hat{p}}}{g/d^{2}} + \frac{\kappa^{N_{0}}_{\hat{p}}(\eta;g,d)}{g/d^{2}} + \left\vert \mathrm{Im} A_{r} \right\vert \cdot \left\vert \frac{(1 + a) \left( 1 - e^{-g^{2}\Lambda_{r}^{2}/d^{2}} \right)}{1 + a\left( 1 - e^{-g^{2}\Lambda_{r}^{2}/d^{2}} \right)} \right\vert, \label{eq:uncert_p_gauss}
\end{gather}
}%
where the systematic uncertainties $\kappa^{N_{0}}_{X}(\eta;g,d)$ are the inverses of
{\small
\begin{gather}
\eta = \sum_{N=1}^{N_0} \mathrm{Bi}\left[N; N_0, r(\phi_{i} \to \phi_{f})\right] \times \max \left[ \left( 1-\frac{\mathrm{Var}_{\hat{x}}(\psi_{f})}{N\kappa^{2}} \right), 0 \right], \\
\eta = \sum_{N=1}^{N_0} \mathrm{Bi}\left[N; N_0, r(\phi_{i} \to \phi_{f})\right] \times \max \left[ \left( 1-\frac{\mathrm{Var}_{\hat{p}}(\psi_{f})}{N\kappa^{2}} \right), 0 \right],
\end{gather}
}%
respectively. From the above two formulae \eqref{eq:uncert_x_gauss} and \eqref{eq:uncert_p_gauss}, one directly verifies that both the statistical and nonlinear terms are dependent on $g$ and $d$ only through the combination $g/d$. Thus, as is mentioned in the Letter, instead of considering the weak limit $g \to 0$, one may equally consider the broad limit of the width $d \to \infty$ to obtain the weak value $A_{w}$ from ${\Delta^{w}_{Q}(g)}/{g}$ and ${\Delta^{w}_{P}(g)}/{(g/d^{2})}$. Another observation is that, by intensifying the interaction $g \to \infty$ while keeping the ratio $g/d$ finite in \eqref{eq:uncert_x_gauss} (or $d \to 0$ while $g/d$ finite for \eqref{eq:uncert_p_gauss}), one can eliminate the contribution from the systematical uncertainty completely. 
In effect, this amounts to amplifying the shifts to infinity while keeping the contributions from the statistical uncertainty and the nonlinearity constant. 

In passing, we note here the correspondent quantities for the conventional indirect projective measurement model under the same assumptions.
Choose any preselection $|\phi_{i}\rangle$ of the system and let $r_{n} := \langle\phi_{i} | E_{n} | \phi_{i}\rangle$ for $A = \lambda_{1}E_{1} + \lambda_{2}E_{2}$.
Based on \eqref{eq:shift_cm} we obtain the shift of the meter as
{\small
\begin{equation}
\Delta_{\hat{x}}^{c}(g,d) = g \cdot \left( E_{\hat{x}}(\phi_{i}) - \Lambda_{m} \right) + g \cdot \Lambda_{m}.
\end{equation}
}%
As for the variance, note that
{\small
\begin{gather}
E_{A}(\phi_{i}) = \lambda_{1}r_{1} + \lambda_{2}r_{2}, \\
E_{A^{2}}(\phi_{i}) = \lambda_{1}^{2}r_{1} + \lambda_{2}^{2}r_{2}.
\end{gather}
}%
Now, observe that
{\small
\begin{align}
E_{A^{2}}(\phi_{i}) &= \lambda_{1}^{2}r_{1} + \lambda_{2}^{2}r_{2} \nonumber \\
    &= \left(\lambda_{1} + \lambda_{2}\right) \left(\lambda_{1}r_{1} + \lambda_{2}r_{2}\right) - \lambda_{1}\lambda_{2}\left(r_{1} + r_{2}\right) \nonumber \\
    &= 2 \Lambda_{m} E_{A}(\phi_{i}) + \left(\Lambda_{r}^{2} - \Lambda_{m}^{2}\right),
\end{align}
}%
which leads to
{\small
\begin{align}
\mathrm{Var}_{A}(\phi_{i}) &:= E_{A^{2}}(\phi_{i}) - E_{A}(\phi_{i})^{2} \nonumber \\
    &= 2 \Lambda_{m} E_{A}(\phi_{i}) + \left(\Lambda_{r}^{2} - \Lambda_{m}^{2}\right) - \left\{ \left(E_{A}(\phi_{i}) - \Lambda_{m}\right)^{2} + 2 \Lambda_{m} E_{A}(\phi_{i}) - \Lambda_{m}^{2}\right\} \nonumber \\
    &= \Lambda_{r}^{2} - \left(E_{A}(\phi_{i}) - \Lambda_{m}\right)^{2}.
\end{align}
}%
Hence from Eq.\eqref{eq:Var_Proj}, one has
{\small
\begin{equation}
\mathrm{Var}_{\mathrm{Id}\otimes\hat{x}}(e^{-igA\otimes P} \phi_{i}\otimes\psi_{i}) = \frac{d^{2}}{2} + g^{2} \cdot \left[ \Lambda_{r}^{2} - \left(E_{A}(\phi_{i}) - \Lambda_{m}\right)^{2} \right].
\end{equation}
}%
As for the momentum part, we just have $\Delta_{\hat{p}}^{c}(g,d) = 0$ and $\mathrm{Var}_{\mathrm{Id}\otimes\hat{p}}(e^{-igA\otimes P} \phi_{i}\otimes\psi_{i}) = 1/(2d^{2})$. Based on formula (10) in the Letter, the total uncertainty for the conventional measurement model can be obtained as
{\small
\begin{equation}
\epsilon^{c}_{\hat{x}}(\eta; g, d) = \frac{\delta_{\hat{x}}}{g} + \sqrt{\frac{d^{2}/(2g^{2}) + \Lambda_{r}^{2} - \left(E_{A}(\phi_{i}) - \Lambda_{m}\right)^{2}}{N_{0}(1-\eta)}}.
\end{equation}
}%

%%%%%%%%%%%%%%%%%%%%%%
\subsection{Discussion}
%%%%%%%%%%%%%%%%%%%%%%

The model discussed in this section is useful for three reasons.  First, the setting is realistic, in the sense that 
the profile of the initial state of the meter is a Gaussian wave function, which is commonly assumed in the actual implementation of experiments.
Second, it has some generality, in the sense that the case in which the physical observable of interest has two eigenvalues is treated. 
Note that most recent experiments concerning weak measurements indeed fall into this category.
Third, the formulae are ready-to-use, in the sense that our formulae for the uncertainties are explicit and can be flexibly adapted to individual situations by tuning the parameters appropriately.

The computational results obtained in this section may be applied both for designing a new experiment or for carefully examining the results obtained by previous weak measurement experiments. In particular, the exact evaluation of the statistical uncertainty obtained in this Letter may be noteworthy in this respect.  In fact, the precise evaluation of the statistical uncertainty
is in general difficult to attain, as the source of the uncertainty involves the probabilistic nature of quantum physics.  To avoid this problem, most
recent experiments concerning weak measurements use laser beams as the composite states of the system and meter, so that the number $N_{0}$ of prepared samples can be treated virtually as infinite, allowing us to neglect the statistical uncertainty.   Now that the exact formulae for evaluating the contribution from statistical uncertainty is obtained, instead of being confined to such experiments by laser beams, for instance, one can conduct interesting experiments using finite, albeit sufficiently large, number of samples.